\documentclass[reqno,12pt,a4paper]{amsart}

\usepackage{mathrsfs}
\usepackage{amssymb}
\usepackage{amsfonts}
\usepackage{latexsym}
\usepackage{amsthm}

\usepackage{graphicx}
\def\lb{\label}

\newcommand{\er}[1]{\textrm{(\ref{#1})}}

\begin{document}

%%%%%%%%%% Some definitions %%%%%%%%%%

%%%%%%%% Equations, theorems %%%%%%%%%
\renewcommand{\theequation}{\arabic{section}.\arabic{equation}}
\theoremstyle{plain}
\newtheorem{theorem}{\bf Theorem}[section]
\newtheorem{lemma}[theorem]{\bf Lemma}
\newtheorem{corollary}[theorem]{\bf Corollary}
\newtheorem{proposition}[theorem]{\bf Proposition}
\newtheorem{definition}[theorem]{\bf Definition}
\newtheorem{condition}[theorem]{\bf Condition}
\newtheorem{remark}[theorem]{\it Remark}
%\theoremstyle{remark}
%\newtheorem{remark}[theorem]{\bf Remark}

%%%%% Alphabet %%%%%
\def\a{\alpha}  \def\cA{{\mathcal A}}     \def\bA{{\bf A}}  \def\mA{{\mathscr A}}
\def\b{\beta}   \def\cB{{\mathcal B}}     \def\bB{{\bf B}}  \def\mB{{\mathscr B}}
\def\g{\gamma}  \def\cC{{\mathcal C}}     \def\bC{{\bf C}}  \def\mC{{\mathscr C}}
\def\G{\Gamma}  \def\cD{{\mathcal D}}     \def\bD{{\bf D}}  \def\mD{{\mathscr D}}
\def\d{\delta}  \def\cE{{\mathcal E}}     \def\bE{{\bf E}}  \def\mE{{\mathscr E}}
\def\D{\Delta}  \def\cF{{\mathcal F}}     \def\bF{{\bf F}}  \def\mF{{\mathscr F}}
\def\c{\chi}    \def\cG{{\mathcal G}}     \def\bG{{\bf G}}  \def\mG{{\mathscr G}}
\def\z{\zeta}   \def\cH{{\mathcal H}}     \def\bH{{\bf H}}  \def\mH{{\mathscr H}}
\def\e{\eta}    \def\cI{{\mathcal I}}     \def\bI{{\bf I}}  \def\mI{{\mathscr I}}
\def\p{\psi}    \def\cJ{{\mathcal J}}     \def\bJ{{\bf J}}  \def\mJ{{\mathscr J}}
\def\vT{\Theta} \def\cK{{\mathcal K}}     \def\bK{{\bf K}}  \def\mK{{\mathscr K}}
\def\k{\kappa}  \def\cL{{\mathcal L}}     \def\bL{{\bf L}}  \def\mL{{\mathscr L}}
\def\l{\lambda} \def\cM{{\mathcal M}}     \def\bM{{\bf M}}  \def\mM{{\mathscr M}}
\def\L{\Lambda} \def\cN{{\mathcal N}}     \def\bN{{\bf N}}  \def\mN{{\mathscr N}}
\def\m{\mu}     \def\cO{{\mathcal O}}     \def\bO{{\bf O}}  \def\mO{{\mathscr O}}
\def\n{\nu}     \def\cP{{\mathcal P}}     \def\bP{{\bf P}}  \def\mP{{\mathscr P}}
\def\r{\rho}    \def\cQ{{\mathcal Q}}     \def\bQ{{\bf Q}}  \def\mQ{{\mathscr Q}}
\def\s{\sigma}  \def\cR{{\mathcal R}}     \def\bR{{\bf R}}  \def\mR{{\mathscr R}}
\def\S{\Sigma}  \def\cS{{\mathcal S}}     \def\bS{{\bf S}}  \def\mS{{\mathscr S}}
\def\t{\tau}    \def\cT{{\mathcal T}}     \def\bT{{\bf T}}  \def\mT{{\mathscr T}}
\def\f{\phi}    \def\cU{{\mathcal U}}     \def\bU{{\bf U}}  \def\mU{{\mathscr U}}
\def\F{\Phi}    \def\cV{{\mathcal V}}     \def\bV{{\bf V}}  \def\mV{{\mathscr V}}
\def\P{\Psi}    \def\cW{{\mathcal W}}     \def\bW{{\bf W}}  \def\mW{{\mathscr W}}
\def\o{\omega}  \def\cX{{\mathcal X}}     \def\bX{{\bf X}}  \def\mX{{\mathscr X}}
\def\x{\xi}     \def\cY{{\mathcal Y}}     \def\bY{{\bf Y}}  \def\mY{{\mathscr Y}}
\def\X{\Xi}     \def\cZ{{\mathcal Z}}     \def\bZ{{\bf Z}}  \def\mZ{{\mathscr Z}}
\def\O{\Omega}
\def\th{\theta}

\newcommand{\gA}{\mathfrak{A}}
\newcommand{\gB}{\mathfrak{B}}
\newcommand{\gC}{\mathfrak{C}}
\newcommand{\gD}{\mathfrak{D}}
\newcommand{\gE}{\mathfrak{E}}
\newcommand{\gF}{\mathfrak{F}}
\newcommand{\gG}{\mathfrak{G}}
\newcommand{\gH}{\mathfrak{H}}
\newcommand{\gI}{\mathfrak{I}}
\newcommand{\gJ}{\mathfrak{J}}
\newcommand{\gK}{\mathfrak{K}}
\newcommand{\gL}{\mathfrak{L}}
\newcommand{\gM}{\mathfrak{M}}
\newcommand{\gN}{\mathfrak{N}}
\newcommand{\gO}{\mathfrak{O}}
\newcommand{\gP}{\mathfrak{P}}
\newcommand{\gQ}{\mathfrak{Q}}
\newcommand{\gR}{\mathfrak{R}}
\newcommand{\gS}{\mathfrak{S}}
\newcommand{\gT}{\mathfrak{T}}
\newcommand{\gU}{\mathfrak{U}}
\newcommand{\gV}{\mathfrak{V}}
\newcommand{\gW}{\mathfrak{W}}
\newcommand{\gX}{\mathfrak{X}}
\newcommand{\gY}{\mathfrak{Y}}
\newcommand{\gZ}{\mathfrak{Z}}

\newcommand{\gm}{\mathfrak{m}}
\newcommand{\gn}{\mathfrak{n}}

\def\ve{\varepsilon}   \def\vt{\vartheta}    \def\vp{\varphi}    \def\vk{\varkappa}

\def\Z{{\mathbb Z}}    \def\R{{\mathbb R}}   \def\C{{\mathbb C}}    \def\K{{\mathbb K}}
\def\T{{\mathbb T}}    \def\N{{\mathbb N}}   \def\dD{{\mathbb D}}

%%%%% Arrows %%%%%

\def\la{\leftarrow}              \def\ra{\rightarrow}            \def\Ra{\Rightarrow}
\def\ua{\uparrow}                \def\da{\downarrow}
\def\lra{\leftrightarrow}        \def\Lra{\Leftrightarrow}

%%%%% Typography %%%%%

\def\lt{\biggl}                  \def\rt{\biggr}
\def\ol{\overline}               \def\wt{\widetilde}
\def\no{\noindent}

%%%%% Math signs %%%%%

\let\ge\geqslant                 \let\le\leqslant
\def\lan{\langle}                \def\ran{\rangle}
\def\/{\over}                    \def\iy{\infty}
\def\sm{\setminus}               \def\es{\emptyset}
\def\ss{\subset}                 \def\ts{\times}
\def\pa{\partial}                \def\os{\oplus}
\def\om{\ominus}                 \def\ev{\equiv}
\def\iint{\int\!\!\!\int}        \def\iintt{\mathop{\int\!\!\int\!\!\dots\!\!\int}\limits}
\def\el2{\ell^{\,2}}             \def\1{1\!\!1}
\def\sh{\sharp}
\def\wh{\widehat}
\def\bs{\backslash}
%%%%% Math operations %%%%%

\def\sh{\mathop{\mathrm{sh}}\nolimits}
\def\Area{\mathop{\mathrm{Area}}\nolimits}
\def\arg{\mathop{\mathrm{arg}}\nolimits}
\def\const{\mathop{\mathrm{const}}\nolimits}
\def\det{\mathop{\mathrm{det}}\nolimits}
\def\diag{\mathop{\mathrm{diag}}\nolimits}
\def\diam{\mathop{\mathrm{diam}}\nolimits}
\def\dim{\mathop{\mathrm{dim}}\nolimits}
\def\dist{\mathop{\mathrm{dist}}\nolimits}
\def\Im{\mathop{\mathrm{Im}}\nolimits}
\def\Iso{\mathop{\mathrm{Iso}}\nolimits}
\def\Ker{\mathop{\mathrm{Ker}}\nolimits}
\def\Lip{\mathop{\mathrm{Lip}}\nolimits}
\def\rank{\mathop{\mathrm{rank}}\limits}
\def\Ran{\mathop{\mathrm{Ran}}\nolimits}
\def\Re{\mathop{\mathrm{Re}}\nolimits}
\def\Res{\mathop{\mathrm{Res}}\nolimits}
\def\res{\mathop{\mathrm{res}}\limits}
\def\sign{\mathop{\mathrm{sign}}\nolimits}
\def\span{\mathop{\mathrm{span}}\nolimits}
\def\supp{\mathop{\mathrm{supp}}\nolimits}
\def\Tr{\mathop{\mathrm{Tr}}\nolimits}
\def\BBox{\hspace{1mm}\vrule height6pt width5.5pt depth0pt \hspace{6pt}}
\def\as{\text{as}}
\def\all{\text{all}}
\def\where{\text{where}}
\def\Dom{\mathop{\mathrm{Dom}}\nolimits}

%%%%%%%%%%%%% specialities %%%%%%%%%%%%%%

\newcommand\nh[2]{\widehat{#1}\vphantom{#1}^{(#2)}}
%{{\mathop{#1}\limits^\wedge}\vphantom{#1}^{(#2)}}
\def\dia{\diamond}

\def\Oplus{\bigoplus\nolimits}

%%%%%%%%%%% End of definitions %%%%%%%%%%

%%%%% OLD OLD OLD

\def\qqq{\qquad}
\def\qq{\quad}
\let\ge\geqslant
\let\le\leqslant
\let\geq\geqslant
\let\leq\leqslant
\newcommand{\ca}{\begin{cases}}
\newcommand{\ac}{\end{cases}}
\newcommand{\ma}{\begin{pmatrix}}
\newcommand{\am}{\end{pmatrix}}
\renewcommand{\[}{\begin{equation}}
\renewcommand{\]}{\end{equation}}
\def\eq{\begin{equation}}
\def\qe{\end{equation}}
\def\[{\begin{equation}}
\def\bu{\bullet}

\newcommand{\fr}{\frac}
\newcommand{\tf}{\tfrac}

\title[Hill's operators with the potentials analytically dependent on energy]
{Hill's operators with the potentials analytically dependent on energy}

\date{\today}
\author[Andrey Badanin]{Andrey Badanin}
\author[Evgeny Korotyaev]{Evgeny L. Korotyaev}
\address{Saint-Petersburg
State University, Universitetskaya nab. 7/9, St. Petersburg,
199034 Russia,
an.badanin@gmail.com,\  a.badanin@spbu.ru,\
korotyaev@gmail.com,\  e.korotyaev@spbu.ru}


\maketitle

\begin{abstract}
We consider Schr\"odinger operators on the line with potentials
that are periodic with respect to the coordinate variable and real
analytic with respect to the energy variable.
We prove that if the imaginary part
of the potential is bounded in the right half-plane, then the high energy spectrum is real,
and the corresponding asymptotics are determined.
Moreover, the Dirichlet and Neumann problems are considered.
These results are used to analyze the good Boussinesq equation.
\end{abstract}

\section{Introduction and main results}
\setcounter{equation}{0}

\subsection{Introduction}
There are a lot of papers about Schr\"odinger operators with potentials
polynomially dependent on energy, see, e.g., the review in \cite{FLM04}.
We consider the wider class of potentials analytically dependent on energy.
Our motivation is related with the good Boussinesq equation on the circle.
McKean \cite{McK81} reduced the third order operator with periodic coefficients, associated
with the good Boussinesq equation, to the Hill equation
with an energy-dependent potential.
This potential is an analytic function of energy
in the domain $\{\l\in\C:|\l|>R,|\arg\l|<\pi-\d\}$, where $R>0$ is large enough
and $\d>0$ is small enough.
Starting from the famous work of Keldysh \cite{Ke71},
operators with a potential polynomially depending on energy were actively studied.
At the same time, we know very few works where operators with a potential
that is an arbitrary analytic function of the spectral parameter would be considered,
see the review below.

We consider Hill's equation
\[
\lb{2oequd}
-y''+ V(x,\l)y=\l y,
\qqq \l\in\mD,
\]
on the whole line where the potential $V(x,\l)$ is 1-periodic with respect
to $x\in\R$ and real analytic with respect to $\l\in\mD$. Here we assume that
$\mD\ss\C$ is a bounded or unbounded domain having a piecewise smooth
boundary $\pa\mD$.
We study the following spectral problems for this equation:

1) the problem on the whole line,

2) the quasi-periodic problems
on the interval $(0,1)$ including the periodic and antiperiodic problems,

3) the Dirichlet problem
$y(0)=y(1)=0$.

Throughout the text, we assume that the potential $ V$ satisfies:

{\it
\no i) For almost every $x\in\R$ the function $ V(x,\cdot)$ is real
analytic in the domain $\mD$,

\no ii) For each $\l\in \mD$ the function $ V$ is 1-periodic and
$ V(\cdot,\l)\in L^1(\T)$, where $\T=\R/\Z$.
}

Some of our results are true for the domains $\mD$
of a quite general form,
while others require additional restrictions on the type of the domain.
Typically, the appearance of the specific domain $\mD$ is dictated by the specifics
of the problem. For example, in the case of the good Boussinesq equation
we are considering, the domain has
the form of a complex plane cut along curves lying in the vicinity of the negative half-line,
see \cite{McK81} and Fig.~\ref{Figbd}~a.
In any case, as a rule, domains containing a segment of the real axis are of interest,
and then we assume that the potential $V$ is a real analytic function.

\medskip

The problem we are considering arises as a result of the reduction of the spectral problem
for a higher order differential operator to a second order one.
Such a reduction for the third-order differential operator
associated with  the good Boussinesq equation on the circle is carried out in the paper
of McKean \cite{McK81}.
Describe briefly the situation, see the details below in Section~\ref{SectBE}.
The {\it good Boussinesq equation}
\[
\lb{gBe}
p_{tt}=-{1\/3} p_{xxxx}-{4\/3}(p^2)_{xx},\qqq p_t=q_x,
\]
is equivalent to the
Lax equation $\dot L=LA-AL$, where
$A=-\pa^2-{4\/3}p$ and the operator $L$
has the form
$
L=\pa^3+\pa p+p\pa+q.
$
Recall that the corresponding L-operator for
the well-studied Korteweg-de Vries equation is the self-adjoint Schr\"odinger operator.
In contrast to this case, the L-operator
for the good Boussinesq is a non-self adjoint third order operator.
This non-self-adjointness greatly complicates the application
of the inverse problem method, since spectral data become non-real
and are more difficult to control.
In \cite{McK81} McKean reduces the spectral problem for the
operator $L$ to the Schr\"odinger equation with an energy-dependent
potential. The equation obtained by McKean is a special case of
equation \er{2oequd} we are considering. The spectrum of the
2-periodic problem is an invariant set with respect to the
Boussinesq flow. The Dirichlet spectrum parameterizes the solutions
of the Boussinesq equation.
The Dirichlet spectrum for the good Boussinesq was the subject of our
work \cite{BK19}.
In our work \cite{BK11} we made the reduction of the spectral problem
for a fourth-order operator to a second order one.

Note that we study here only the case of the good Boussinesq equation on the circle.
The associated operator $L$ is non-self-adjoint, however,
the high energy spectra for the corresponding Schr\"odinger equation with an energy-dependent
potential localizes near the real axis.
The situation for the bad Boussinesq equation  is completely different.
The associated operator $i\pa^3+i\pa p+ip\pa+q$ is self-adjoint but the high energy
spectra for the corresponding Schr\"odinger equation with an energy-dependent
potential localizes far from the real axis.
We considered this operator in our paper \cite{BK15}.
The spectral properties of higher order differential operators
with periodic coefficients were the subject of
Badanin and Korotyaev \cite{BK11}, \cite{BK12}, Papanicolaou \cite{P95}, \cite{P03},
see also references therein.

Schr\"odinger operators with polynomially energy-dependent potentials are also well studied,
see, e.g., Alonso \cite{A80},
Jaulent and Jean \cite{JJ76}, \cite{JJ76x},
Kamimura \cite{Ka08}, see also the book \cite{Ma12}
and references therein, moreover,
there is enormous physical and technical literature on this subject.
By the well-known technique developed by Keldysh \cite{Ke71},
these problems are reduced to vector
spectral problems where the potential does not depend on the spectral parameter.
We consider a much wider class of problems when the potential
is an arbitrary holomorphic function of the spectral parameter.
Keldysh's approach does not work in this case and
these problems are much worse studied.
In connection with this subject, we mention the papers
McKean \cite{McK81} and Badanin--Korotyaev \cite{BK11} for the periodic problems,
and Calogero--Jagannathan \cite{CJ67} for the scattering problems.
Note that there are a large number of articles where the certain special classes of
holomorphic families of operators with respect to an additional parameter are considered,
see Derkach and Malamud \cite{DM89},
Gesztesy, Kalton, Makarov and Tsekanovskii \cite{GKMT01} and references therein.

\subsection{The definitions}
We analyze equation \er{2oequd} on the whole line using the direct integral decomposition.
In order to describe this decomposition
we introduce  the  operators on $L^2(0,1)$ given by
\[
\lb{defHp}
H(k,\l)=H_o(k)+ V(\cdot,\l),\qq  k\in[0,2\pi),
\]
where $\l$ belongs to the domain $\mD$ and the unperturbed operators $H_o(k)$
have the form $H_o(k)y=-y''$
under the quasi-periodic boundary
conditions
\[
\lb{fbc}
y(1)=e^{ik}y(0),\qqq y'(1)=e^{ik}y'(0),\qqq k\in[0,2\pi).
\]
If $k=0$, then the conditions \er{fbc} are called {\it periodic} conditions,
if $k=\pi$, then they are called {\it antiperiodic} ones,
jointly they are {\it 2-periodic} conditions.

%Each operator $H_o(k),k\in[0,2\pi)$, is unitarily equivalent to
%the operator
%$$
%\wt H_o(k)y=(-i\pa+k)^2y,\qq y(1)=y(0),\qq
%y'(1)=y'(0).
%$$
%The family $\wt H_o(\cdot)$ is analytic
%in the sense of Kato.

Recall the following standard definitions.
Let $k\in[0,2\pi)$.
The point $\l\in\mD$ is called the {\it regular point} of
the operator-valued function $H(k,\l)$,
if the resolvent
$(H(k,\l)-\l)^{-1}
$
exists and bounded. We denote by $\rho(H(k,\cdot))$ the  set of all regular
points of the operator-valued function $H(k,\l)$. The operator-valued function
$(H(k,\l)-\l)^{-1}$ is analytic on the set $\rho(H(k,\cdot))$.
The {\it spectrum} $\s(H(k,\cdot))$ of the function $H(k,\l)$ is the set
$$
\s(H(k,\cdot))=\mD\sm\rho(H(k,\cdot)).
$$
The  set $\s(H(k,\cdot))$ is closed.
The number $\l_o\in\mD$ is called the {\it eigenvalue} of the operator-valued function
$H(k,\l)$,
if the equation
$$
H(k,\l_o)y_o=\l_oy_o
$$
has a non-trivial solution, the corresponding solution
$y_o$ is called the {\it eigenvector}.
The spectrum $\s(H_o(0))\cup\s(H_o(\pi))$ of the 2-periodic problem
for the unperturbed operator $H_o$ is pure discrete,
consists of the simple eigenvalue $\l_0^{o,+}=0$ and the eigenvalues
$ \l_{n}^{o,\pm}=(\pi n)^2, n\in\N$, of multiplicity 2.
We show in Theorem~\ref{ThSpec} that the spectrum in the perturbed case is also discrete.

Moreover, introduce the operator-valued function
\[
\lb{defHd}
T(\l)=T_o+ V(\cdot,\l)
\]
in the domain $\mD$, where the unperturbed operator $T_o$ in $L^2(0,1)$
has the form $T_oy=-y''$ with the Dirichlet boundary conditions
\[
\lb{dbc}
y(0)=y(1)=0.
\]
The point $\l\in\mD$ is a {\it regular point} of the function $T(\l)$,
if the resolvent
$
%\cR_d(\l)=
(T(\l)-\l)^{-1}
$
exists and bounded. The operator-valued function
$(T(\l)-\l)^{-1}$ is analytic on the set $\rho(T)$ of  all regular
points of the operator-valued function $H(k,\l)$.
The {\it spectrum} $\s(T)$ is the set
$$
\s(T)=\mD\sm\rho(T).
$$
The Dirichlet spectrum $\s(T_o)$
for the unperturbed operator $T_o$  consists of the
simple eigenvalues $\gm_n^o=(\pi n)^2,n\in\N$.

Similarly, we define the operator $\cN(\l)$ of the Neumann problem by
\[
\lb{defHn}
\cN(\l)=\cN_o+ V(\cdot,\l),\qqq\l\in\mD,
\]
where the unperturbed operator $\cN_oy=-y''$ acts on the functions $y$ such that
\[
\lb{nbc}
y'(0)=y'(1)=0.
\]
Let us denote by $\s(\cN)$ the spectrum of the operator $\cN$.
The spectrum $\s(\cN_o)$ for the unperturbed operator consists of the
simple eigenvalues $\gn_n^o=(\pi n)^2,n=0,1,2,...$

Introduce the operators $H(\l),\l\in\mD$, acting on $L^2(\R)$, by
\[
\lb{defHw}
H(\l)=H_o+ V(\cdot,\l),
\]
where the unperturbed operator $H_o$ in $L^2(\R)$
has the form
$$
H_o y=-y''.
$$
Now we write  the direct integral decomposition for the operator-valued function $H(\l)$.
Introduce the Hilbert spaces
\[
\lb{intH}
\mH'=L^2([0,1],dt),\qqq \mH=\int_{[0,2\pi)}^\os \mH' \ {d
k\/2\pi}
\]
Introduce the unitary operator $U:L^2(\R)\to\mH$
by
\[
\lb{gt}
(U f)_k(t)=\sum_{n\in\Z}e^{-in k}f(t+n),\qqq
(k,t)\in[0,2\pi)\ts[0,1].
\]
Now we formulate our preliminary results about
the direct integral decomposition of the operator-valued functions $H(\l)$
given by \er{defHw}.

\begin{proposition}
\lb{ThSpec}
i) The operator-valued function $H(\l)$ satisfies
\[
\lb{deH}
UH(\l)U^{-1}=\int_{[0,2\pi)}^\os H( k,\l){d k\/2\pi},\qqq\l\in\mD,
\]
where $U$ is defined by \er{gt}.

ii) The spectra $\s(H(k,\cdot))$ for each $k\in[0,2\pi)$, $\s(T)$ and $\s(\cN)$ are pure discrete.

iii) Each eigenvalue $\l(k)\in\mD$ of  the operator $H(k,\cdot)$ is
a piecewise analytic and $2\pi$-periodic function of $k\in\R$. Moreover,
$\s(H(2\pi-k,\cdot))=\s(H(k,\cdot))$ for all $k\in[0,2\pi)$,
counting with multiplicities.

iv) The spectrum $\s(H)$ of the operator-valued function $H(\l)$ satisfies
\[
\lb{specH}
\s(H)=\ol{\cup_{k\in[0,\pi]}\s(H(k,\cdot))}.
\]

\end{proposition}

\no {\bf Remark.}
1) The spectrum $\s(H_o)$ of the unperturbed operator $H_o$
on the whole line is pure
absolutely continuous, has multiplicity 2, and satisfies
$\s(H_o)=[0,+\iy)$.

2) We consider the band functions $\l_n(k),k\in[0,2\pi)$
mainly for high energy.
Note that if the eigenvalue $\l_n(k)$ goes to the boundary of the domain $\mD$,
then it leaves the spectrum of $H(k,\l)$ and, therefore,
does not generate the spectrum of $H(\l)$.

\subsection{Main results}
Introduce the notations
$$
\l=\mu+i\nu,\qq
Q=\Im V,
$$
and the norm of the potential
$$
\|V(\cdot,\l)\|=\int_0^1|V(x,\l)|dx,\qqq
% \|Q(\cdot,\l)\|_1=\sup_{x\in[0,1]}|Q(x,\l)|,\qqq
\l\in\mD.
$$
Now we formulate our first results about the spectra.

\begin{theorem}
\lb{ThSpec2}
Let $I\ss\R$ be a finite or infinite interval, $I\ss\ol\mD$,
let for a.e. $x\in\R$ the function  $Q=\Im V$ satisfy
$$
Q(x,\cdot)\in C(\ol\mD),\qq {\pa Q\/\pa\nu}(x,\cdot)\in C(\ol\mD),
$$
and
\[
\lb{locestdil}
\sup_{(x,\l)\in [0,1]\ts I}\Big|{\pa Q(x,\l)\/\pa\nu}\Big|<1.
\]
Then the spectral set $\gS$, defined by
\[
\lb{spectra}
\gS=\s(H)\cup\s(T)\cup\s(\cN),
\]
for some $\d>0$ satisfies
\[
\lb{locrsp}
\gS\cap\big(I\ts (-\d,\d)\big)\cap(\mD\cup I)\ss I.
\]
\end{theorem}

\medskip

Thus, the estimate \er{locestdil} guarantees that the spectrum
is real in the vicinity of the real axis. In the following Theorem
we obtain the conditions when the spectrum in a half-plane is real.
Introduce the domains
$$
\Pi_a=\{\l\in\C:\Re\l>a\},\qqq
\mD_a=\mD\cap\Pi_a, \qqq a\in\R,
$$
and for a domain $\O\ss\mD$ we introduce the functional
\[
\lb{defunc}
\xi(\O)=\sup_{(x,\l)\in[0,1]\ts\O}|Q(x,\l)|.
\]
Theorem~\ref{Thspec} gives that if $Q$ is bounded on the right half-plane,
then the high energy spectra in this half-plane is real.

\begin{theorem}
\lb{Thspec}
Let the potential $V$ satisfy the estimate
\[
\lb{cestimV}
\xi(\mD_{a})<\iy
\]
for some $a\in\R$.
Let, in addition,
\[
\lb{cestimV2}
(a,+\iy)\ts(-\rho,\rho)\ss\mD,\qq \where\qq
\rho={\xi(\mD_{a})\/2-\sqrt3}.
\]
Then the spectra $\s(H)$, $\s(T)$ and $\s(\cN)$ in the domain $\mD_{a+ \rho}$ are real:
\[
\lb{locsphs}
\gS\cap\mD_{a+ \rho}\ss(a+ \rho,+\iy),
\]
where $\gS=\s(H)\cup\s(T)\cup\s(\cN)$.
In particular, if the half-plane $\Pi_a\ss\mD$
and $\xi(\Pi_{a})<\iy$,
then the spectra in the half-plane $\Pi_{a+\rho_1}$ are real:
\[
\lb{locsphs1}
\gS\cap\Pi_{a+ \rho_1}\ss(a+ \rho_1,+\iy),\qqq\rho_1={\xi(\Pi_{a})\/2-\sqrt3}.
\]

\end{theorem}

\no{\bf Remark.} 1) The conditions of Theorem~\ref{Thspec} are more restrictive
than the condition \er{locestdil} of Theorem~\ref{ThSpec2} in the following sense.
Assume that the restrictions of Theorem~\ref{Thspec} hold true,
that is assume that $\xi(\mD_{a})<\iy$ for some $a\in\R$ and $(a,+\iy)\ts(-\rho,\rho)\ss\mD$,
where $\rho$ is given by \er{cestimV2}.
Then the estimate \er{estda} of Lemma~\ref{Lmehf} gives
$$
\sup_{(x,\l)\in[0,1]\ts(a+\rho,+\iy)}\Big|{\pa Q(x,\l)\/\pa\nu}\Big|
\le{2\/\rho} \xi(\mD_{a})
=2(2-\sqrt3)<1,
$$
that is the restriction \er{locestdil} is fulfilled for the interval $I=(a+\rho,+\iy)$.

2) This is an open question:
is it possible to take the constant in the definitions \er{cestimV2}
and \er{locsphs1} more than $2-\sqrt3$?

3) We illustrate Theorems~\ref{ThSpec2} and \ref{Thspec}
with two simple examples in Section~\ref{SSex}.

\subsection{High energy asymptotics}
Theorem \ref{ThSpec2} provides that if $(a,+\iy)\ts(-r,r)\ss\mD$
for some $(a,r)\in\R\ts\R_+$ and
the potential $ V$ satisfies the condition
\[
\lb{estdQhl}
\sup_{(x,\l)\in [0,1]\ts (a,+\iy)}\Big|{\pa Q(x,\l)\/\pa\nu}\Big|<1,
\]
then the spectra in the half-strip $(a,+\iy)\ts(-\d,\d)$ for some $\d>0$ are real.
In the following theorem we show that the high energy spectra in this case
are similar to the spectra for the standard Hill operator
with the real potential which does not depend on energy and determine
high energy asymptotics of the eigenvalues.

\begin{theorem}
\lb{Thasspec}
Let $(a,+\iy)\ts(-r,r)\ss\mD$ for some $(a, r)\in\R\ts\R_+$.
Let the potential $ V$ satisfy the condition \er{estdQhl}
and let $b\ge a$ be large enough.

i) Let, in addition, $\| V(\cdot,\l)\|=\l^{1\/2}o(1)$
as $\l\to+\iy$.
 Then the eigenvalues $\l_{2n}^\pm\in\mD_b$ of the operator $H(0,\l)$
and the eigenvalues  $\l_{2n-1}^\pm\in\mD_b$ of the operator $H(\pi,\l)$ are real and satisfy
\[
\lb{indp1}
\l_{N-1}^+<\l_{N}^-\le\l_{N}^+<\l_{N+1}^-\le\l_{N+1}^+<...
\]
for some $N\in\N$.
The spectrum $\s(H)$ in the domain $\mD_b$ is real, absolutely continuous,
has multiplicity two, and consists of the intervals $[\l_{n-1}^+,\l_n^-],n\ge N$,
separated by the gaps $(\l_n^-,\l_n^+)$
\[
\lb{specHreal}
\s(H)\cap\mD_b=\cup_{n\ge N}[\l_{n-1}^+,\l_n^-]\ss\R.
\]
The eigenvalues $\gm_n\in\mD_b$ of the Dirichlet operator
$T(\l)$ are real, simple and satisfy
\[
\lb{dsing}
\gm_N<\gm_{N+1}<\gm_{N+2}<...,\qq
\gm_n\in[\l_{n}^-,\l_{n}^+],\qq n=N,N+1,...,
\]
and there are no other Dirichlet eigenvalues in $\mD_{b}$.
The eigenvalues $\gn_n\in\mD_b$ of the Neumann operator
$\cN(\l)$ are real, simple and satisfy
\[
\lb{nsing}
\gn_N<\gn_{N+1}<\gn_{N+2}<...,\qq
\gn_n\in[\l_{n}^-,\l_{n}^+],\qq n=N,N+1,...,
\]
and there are no other Neumann eigenvalues in $\mD_{b}$.

ii) Let, in addition, $\| V(\cdot,\l)\|=\l^{1\/6}o(1)$ as $\l\to+\iy$, and let
\[
\lb{condVav}
\int_0^1 V(s,\l)ds=o(1),
\]
as $|\l|\to\iy,\l\in\mD_{a}$.
Then the eigenvalues of the 2-periodic problem satisfy
\[
\lb{as2p}
\l_{n}^\pm=(\pi n)^2+o(1)\qq\as\qq n\to+\iy.
\]

\end{theorem}

\no {\bf Remark.}
1) The results, similar to \er{dsing},
for the Neumann eigenvalues hold, see Remark after Lemma~\ref{Lm2pev}.

2) The condition \er{condVav} can be written
in the slightly more general form
$\wh V_o(\l)=C+o(1)$
for some $C\in\R$ independent of $\l$
but the constant $C$ is removed by shifting the spectral parameter.

3) Korotyaev \cite{K99} determined
the sharp spectral asymptotics for Schr\"odinger operators with periodic complex
potentials.

\medskip

The plan of the paper is as follows. We discuss the relations between our second order
operator and the third order operator associated with the good Bossinesq equation.
 In Section 3 we calculate the resolvent
and prove that the spectra $\s(H(k,\cdot)),k\in[0,2\pi)$, as well as the spectrum $\s(T)$,
are sets of zeros of certain functions analytic in the domain $\mD$.
It follows that the spectra $\s(H(k,\cdot))$ and $\s(T)$ are discrete
and we obtain their description in terms of zeros of the analytic functions.
Moreover, in Section 4 we prove Proposition~\ref{ThSpec}
on the direct integral decomposition for the operator $H(\l)$.
In Section 5 we establish the conditions when the spectrum is real
and prove Theorems~\ref{ThSpec2} and \ref{Thspec}.
In addition, we consider two simple examples there.
In Section 6 we study high energy asymptotic behavior of the spectra
and prove Theorem~\ref{Thasspec}.
Moreover, there we prove Corollary~\ref{Corbe} for the good Boussinesq equation.

\section{Relationship with the good Boussinesq equation}
\setcounter{equation}{0}
\lb{SectBE}

\subsection{Ramifications and three-point eigenvalues}
Recall that the good Boussinesq equation \er{gBe}
is equivalent to the
Lax equation $\dot L=LA-AL$, where
the non-self-adjoint operator $L$, acting on $L^2(\R)$,
has the form
$$
L=\pa^3+\pa p+p\pa+q.
$$
We consider the operator $L$ in the class of real
1-periodic coefficients $p',q\in L^1(\T)$.
The operator $L$ with smooth coefficients $p,q$ was studied by McKean \cite{McK81}.
The following results from  \cite{McK81} can be extended from the class
of the  smooth coefficients onto the class  $p',q\in L^1(\T)$.

Introduce the fundamental solutions $y_j(x,\zeta),j=1,2,3$, of the equation
\[
\lb{1b}
y'''+(py)'+py'+q y=\zeta y,\qqq (x,\zeta)\in\R\ts\C,
\]
satisfying the conditions $y_j^{(k-1)}(0,\zeta)=\d_{jk}$.
Let $M(x,\zeta)$ be the matrix $M=(y_j^{(k-1)})_{j,k=1}^3$,
$M(0,\zeta)=\1_3$ is a $3\ts3$-identical matrix.
Each matrix-valued function $M(x,\cdot),x\in\R$, is entire.
The matrix $M(1,\zeta)$ is the {\it monodromy matrix}. The eigenvalues
$\vk_j,j=1,2,3$, of the monodromy matrix are the {\it multipliers},
they satisfy the identity $\vk_1\vk_2\vk_3=1$.
The functions $\vk_j=\vk_j(\zeta)$ constitute three branches
of the function, analytic on a 3-sheeted multiplier Riemann surface $\mR$, see \cite{McK81}
(the similar surface for the bad Boussinesq is described in \cite{BK15}).
{\it Ramifications} of this surface
are points where two or all three functions take the same value.
They are the zeros of the entire function
$(\vk_1-\vk_2)^2(\vk_1-\vk_3)^2(\vk_2-\vk_3)^2$ called the {\it discriminant},
see \cite{McK81}, \cite{BK14} and \cite{BK15}.
There are a finite number of the ramifications in any bounded domain in $\C$.
The set $\{r_n^\pm\}_{n\in\Z}$ of ramifications is invariant
with respect to the Boussinesq flow.

To each multiplier $\vk_j(\zeta),j=1,2,3,$ corresponds the Floquet solution
$\p_j(x,\zeta),(x,\zeta)\in\R\ts\C$,
satisfying the conditions
$$
\p_j(0,\zeta)=1,\qqq\p_j(x+1,\zeta)=\vk_j\p_j(x,\zeta).
$$
For each $x\in\R$ the functions $\p_j(x,\cdot)$ constitute three branches
of the function, meromorphic  on the surface $\mR$.
The {\it set of poles} of the functions $\p_j(x,\cdot)$ coincides with the spectrum
$\{\zeta_n\}_{n\in\Z\sm\{0\}}$
of the three-point Dirichlet problem
$$
y'''+(py)'+py'+q y=\zeta y,\qqq
y(0)=y(1)=y(2).
$$
This problem was the subject of our paper \cite{BK19}.

In the unperturbed case $p=q=0$ the ramifications $r_n^{0,\pm},n\in\Z$, and the three-point
eigenvalues $\zeta_n^0,n\in\Z\sm\{0\}$,
have the form $r_n^{0,\pm}=\zeta_n^0=({2\pi n\/\sqrt3})^3,n\in\Z\sm\{0\},
r_0^{0,\pm}=0$, see \cite{McK81}.
In the perturbed case the sets of the ramifications and of the three-point Dirichlet eigenvalues
are symmetric with respect to the real line.
Moreover, the three-point eigenvalues at high energy are real and simple and satisfy
\cite{BK19}
$$
\zeta_n=\Big({2\pi n\/\sqrt3}\Big)^3-{4\pi n\/\sqrt3} p_0
+{2\pi n\/\sqrt3} \wt p_n+q_0-\wt q_n +O(n^{-{1\/2}}),
$$
as $n\to\pm\iy$, where
$$
\wt f_n=
{2\/\sqrt3}\int_0^1 f(x)\cos \Big(2\pi nx+{\pi\/6}\Big)dx,\qq n\in\N.
$$

\begin{figure}
\tiny
\unitlength 0.6mm % = 2.845pt
\linethickness{0.4pt}
\ifx\plotpoint\undefined\newsavebox{\plotpoint}\fi % GNUPLOT compatibility
\begin{picture}(229,59.75)(8,35)
\put(70.25,94.75){\line(0,-1){53.75}}
\put(187.25,94.75){\line(0,-1){53.75}}
\put(20.5,65.5){\line(1,0){105.75}}
\put(237,65.5){\line(-1,0){105.75}}
%\emline(20.75,73.25)(70,65.75)
\multiput(20.75,73.25)(.220852018,-.033632287){223}{\line(1,0){.220852018}}
%\end
%\emline(236.75,73.25)(187.5,65.75)
\multiput(236.75,73.25)(-.220852018,-.033632287){223}{\line(-1,0){.220852018}}
%\end
%\emline(20.75,58)(70,65.5)
\multiput(20.75,58)(.220852018,.033632287){223}{\line(1,0){.220852018}}
%\end
%\emline(236.75,58)(187.5,65.5)
\multiput(236.75,58)(-.220852018,.033632287){223}{\line(-1,0){.220852018}}
%\end
\qbezier(42.5,70.25)(42.25,68)(42,65.75)
\qbezier(215,70.25)(215.25,68)(215.5,65.75)
\put(42.5,73){\makebox(0,0)[cc]{$\d$}}
\put(215,73){\makebox(0,0)[]{$\d$}}
\put(96.5,91.25){\makebox(0,0)[cc]{$a)$}}
\put(161,91.25){\makebox(0,0)[]{$b)$}}
%\circle(70.25,65){38.275}
\put(89.388,65){\line(0,1){.8952}}
\put(89.367,65.895){\line(0,1){.8932}}
\put(89.304,66.788){\line(0,1){.8893}}
\multiput(89.199,67.678)(-.029199,.176689){5}{\line(0,1){.176689}}
\multiput(89.053,68.561)(-.031193,.145941){6}{\line(0,1){.145941}}
\multiput(88.866,69.437)(-.032559,.123705){7}{\line(0,1){.123705}}
\multiput(88.638,70.303)(-.033521,.106791){8}{\line(0,1){.106791}}
\multiput(88.37,71.157)(-.030784,.084085){10}{\line(0,1){.084085}}
\multiput(88.062,71.998)(-.03153,.075048){11}{\line(0,1){.075048}}
\multiput(87.715,72.823)(-.032089,.067367){12}{\line(0,1){.067367}}
\multiput(87.33,73.632)(-.0324971,.060731){13}{\line(0,1){.060731}}
\multiput(86.908,74.421)(-.0327807,.0549198){14}{\line(0,1){.0549198}}
\multiput(86.449,75.19)(-.0329595,.0497713){15}{\line(0,1){.0497713}}
\multiput(85.954,75.937)(-.0330483,.0451641){16}{\line(0,1){.0451641}}
\multiput(85.426,76.659)(-.0330586,.0410059){17}{\line(0,1){.0410059}}
\multiput(84.864,77.357)(-.0329994,.037225){18}{\line(0,1){.037225}}
\multiput(84.27,78.027)(-.032878,.0337648){19}{\line(0,1){.0337648}}
\multiput(83.645,78.668)(-.0344214,.03219){19}{\line(-1,0){.0344214}}
\multiput(82.991,79.28)(-.0378833,.0322415){18}{\line(-1,0){.0378833}}
\multiput(82.309,79.86)(-.0416647,.0322245){17}{\line(-1,0){.0416647}}
\multiput(81.601,80.408)(-.0458218,.0321303){16}{\line(-1,0){.0458218}}
\multiput(80.868,80.922)(-.0504262,.0319485){15}{\line(-1,0){.0504262}}
\multiput(80.111,81.401)(-.0555701,.0316658){14}{\line(-1,0){.0555701}}
\multiput(79.333,81.845)(-.0613744,.031265){13}{\line(-1,0){.0613744}}
\multiput(78.535,82.251)(-.074182,.033516){11}{\line(-1,0){.074182}}
\multiput(77.719,82.62)(-.083236,.033011){10}{\line(-1,0){.083236}}
\multiput(76.887,82.95)(-.094099,.032312){9}{\line(-1,0){.094099}}
\multiput(76.04,83.241)(-.107446,.03136){8}{\line(-1,0){.107446}}
\multiput(75.181,83.491)(-.124337,.030056){7}{\line(-1,0){.124337}}
\multiput(74.31,83.702)(-.146541,.028242){6}{\line(-1,0){.146541}}
\multiput(73.431,83.871)(-.22155,.03203){4}{\line(-1,0){.22155}}
\put(72.545,83.999){\line(-1,0){.8912}}
\put(71.654,84.086){\line(-1,0){.8943}}
\put(70.759,84.131){\line(-1,0){.8954}}
\put(69.864,84.134){\line(-1,0){.8946}}
\put(68.969,84.095){\line(-1,0){.8918}}
\multiput(68.077,84.014)(-.22175,-.03061){4}{\line(-1,0){.22175}}
\multiput(67.19,83.891)(-.176064,-.032758){5}{\line(-1,0){.176064}}
\multiput(66.31,83.728)(-.124528,-.029256){7}{\line(-1,0){.124528}}
\multiput(65.438,83.523)(-.107645,-.030668){8}{\line(-1,0){.107645}}
\multiput(64.577,83.277)(-.094305,-.031706){9}{\line(-1,0){.094305}}
\multiput(63.729,82.992)(-.083446,-.032474){10}{\line(-1,0){.083446}}
\multiput(62.894,82.667)(-.074396,-.033038){11}{\line(-1,0){.074396}}
\multiput(62.076,82.304)(-.066705,-.033442){12}{\line(-1,0){.066705}}
\multiput(61.275,81.903)(-.0600629,-.0337159){13}{\line(-1,0){.0600629}}
\multiput(60.494,81.464)(-.0506307,-.0316234){15}{\line(-1,0){.0506307}}
\multiput(59.735,80.99)(-.0460276,-.0318348){16}{\line(-1,0){.0460276}}
\multiput(58.999,80.481)(-.0418711,-.0319557){17}{\line(-1,0){.0418711}}
\multiput(58.287,79.937)(-.0380899,-.0319971){18}{\line(-1,0){.0380899}}
\multiput(57.601,79.361)(-.0346278,-.0319678){19}{\line(-1,0){.0346278}}
\multiput(56.943,78.754)(-.0330945,-.0335526){19}{\line(0,-1){.0335526}}
\multiput(56.314,78.117)(-.0332382,-.0370119){18}{\line(0,-1){.0370119}}
\multiput(55.716,77.45)(-.0333218,-.0407924){17}{\line(0,-1){.0407924}}
\multiput(55.15,76.757)(-.0333382,-.0449506){16}{\line(0,-1){.0449506}}
\multiput(54.616,76.038)(-.0332791,-.0495582){15}{\line(0,-1){.0495582}}
\multiput(54.117,75.294)(-.0331334,-.0547078){14}{\line(0,-1){.0547078}}
\multiput(53.653,74.528)(-.0328872,-.0605207){13}{\line(0,-1){.0605207}}
\multiput(53.226,73.742)(-.032522,-.067159){12}{\line(0,-1){.067159}}
\multiput(52.835,72.936)(-.032013,-.074844){11}{\line(0,-1){.074844}}
\multiput(52.483,72.112)(-.031324,-.083885){10}{\line(0,-1){.083885}}
\multiput(52.17,71.274)(-.030407,-.094732){9}{\line(0,-1){.094732}}
\multiput(51.896,70.421)(-.033355,-.123493){7}{\line(0,-1){.123493}}
\multiput(51.663,69.556)(-.032132,-.145738){6}{\line(0,-1){.145738}}
\multiput(51.47,68.682)(-.030335,-.176497){5}{\line(0,-1){.176497}}
\put(51.318,67.8){\line(0,-1){.8886}}
\put(51.208,66.911){\line(0,-1){.8928}}
\put(51.14,66.018){\line(0,-1){.895}}
\put(51.113,65.123){\line(0,-1){.8953}}
\put(51.128,64.228){\line(0,-1){.8936}}
\put(51.185,63.334){\line(0,-1){.89}}
\multiput(51.284,62.444)(.028062,-.176873){5}{\line(0,-1){.176873}}
\multiput(51.424,61.56)(.030254,-.146139){6}{\line(0,-1){.146139}}
\multiput(51.606,60.683)(.031763,-.123912){7}{\line(0,-1){.123912}}
\multiput(51.828,59.816)(.032834,-.107004){8}{\line(0,-1){.107004}}
\multiput(52.091,58.96)(.033603,-.093646){9}{\line(0,-1){.093646}}
\multiput(52.393,58.117)(.031047,-.075249){11}{\line(0,-1){.075249}}
\multiput(52.735,57.289)(.031655,-.067572){12}{\line(0,-1){.067572}}
\multiput(53.115,56.478)(.0321057,-.0609389){13}{\line(0,-1){.0609389}}
\multiput(53.532,55.686)(.0324267,-.0551296){14}{\line(0,-1){.0551296}}
\multiput(53.986,54.914)(.0326386,-.0499823){15}{\line(0,-1){.0499823}}
\multiput(54.475,54.164)(.0327571,-.0453758){16}{\line(0,-1){.0453758}}
\multiput(55,53.438)(.0327941,-.0412178){17}{\line(0,-1){.0412178}}
\multiput(55.557,52.738)(.0327592,-.0374365){18}{\line(0,-1){.0374365}}
\multiput(56.147,52.064)(.03266,-.0339757){19}{\line(0,-1){.0339757}}
\multiput(56.767,51.418)(.0342136,-.0324108){19}{\line(1,0){.0342136}}
\multiput(57.417,50.803)(.0376751,-.0324846){18}{\line(1,0){.0376751}}
\multiput(58.096,50.218)(.0414565,-.0324919){17}{\line(1,0){.0414565}}
\multiput(58.8,49.665)(.0456141,-.0324244){16}{\line(1,0){.0456141}}
\multiput(59.53,49.147)(.0502196,-.0322723){15}{\line(1,0){.0502196}}
\multiput(60.283,48.663)(.0553652,-.0320227){14}{\line(1,0){.0553652}}
\multiput(61.059,48.214)(.061172,-.0316593){13}{\line(1,0){.061172}}
\multiput(61.854,47.803)(.067801,-.03116){12}{\line(1,0){.067801}}
\multiput(62.667,47.429)(.083022,-.033545){10}{\line(1,0){.083022}}
\multiput(63.498,47.093)(.093889,-.032917){9}{\line(1,0){.093889}}
\multiput(64.343,46.797)(.107242,-.03205){8}{\line(1,0){.107242}}
\multiput(65.201,46.541)(.124141,-.030856){7}{\line(1,0){.124141}}
\multiput(66.069,46.325)(.146356,-.029184){6}{\line(1,0){.146356}}
\multiput(66.948,46.15)(.22134,-.03346){4}{\line(1,0){.22134}}
\put(67.833,46.016){\line(1,0){.8907}}
\put(68.724,45.923){\line(1,0){.894}}
\put(69.618,45.873){\line(1,0){.8954}}
\put(70.513,45.864){\line(1,0){.8948}}
\put(71.408,45.898){\line(1,0){.8923}}
\put(72.3,45.973){\line(1,0){.8878}}
\multiput(73.188,46.089)(.176271,.031625){5}{\line(1,0){.176271}}
\multiput(74.069,46.247)(.145499,.033196){6}{\line(1,0){.145499}}
\multiput(74.942,46.447)(.10784,.029974){8}{\line(1,0){.10784}}
\multiput(75.805,46.686)(.094507,.031099){9}{\line(1,0){.094507}}
\multiput(76.656,46.966)(.083654,.031937){10}{\line(1,0){.083654}}
\multiput(77.492,47.286)(.074607,.032559){11}{\line(1,0){.074607}}
\multiput(78.313,47.644)(.066919,.033012){12}{\line(1,0){.066919}}
\multiput(79.116,48.04)(.0602786,.0333288){13}{\line(1,0){.0602786}}
\multiput(79.899,48.473)(.0544641,.0335325){14}{\line(1,0){.0544641}}
\multiput(80.662,48.943)(.0493135,.0336405){15}{\line(1,0){.0493135}}
\multiput(81.402,49.447)(.0447056,.033666){16}{\line(1,0){.0447056}}
\multiput(82.117,49.986)(.0405477,.0336191){17}{\line(1,0){.0405477}}
\multiput(82.806,50.558)(.0367679,.0335079){18}{\line(1,0){.0367679}}
\multiput(83.468,51.161)(.0333097,.033339){19}{\line(0,1){.033339}}
\multiput(84.101,51.794)(.0334757,.0367973){18}{\line(0,1){.0367973}}
\multiput(84.703,52.456)(.0335835,.0405772){17}{\line(0,1){.0405772}}
\multiput(85.274,53.146)(.0336268,.0447351){16}{\line(0,1){.0447351}}
\multiput(85.812,53.862)(.0335972,.049343){15}{\line(0,1){.049343}}
\multiput(86.316,54.602)(.0334847,.0544935){14}{\line(0,1){.0544935}}
\multiput(86.785,55.365)(.0332759,.0603078){13}{\line(0,1){.0603078}}
\multiput(87.218,56.149)(.032953,.066948){12}{\line(0,1){.066948}}
\multiput(87.613,56.952)(.032493,.074636){11}{\line(0,1){.074636}}
\multiput(87.971,57.773)(.031863,.083682){10}{\line(0,1){.083682}}
\multiput(88.289,58.61)(.031016,.094534){9}{\line(0,1){.094534}}
\multiput(88.568,59.461)(.02988,.107866){8}{\line(0,1){.107866}}
\multiput(88.807,60.324)(.033069,.145528){6}{\line(0,1){.145528}}
\multiput(89.006,61.197)(.03147,.176299){5}{\line(0,1){.176299}}
\put(89.163,62.079){\line(0,1){.8879}}
\put(89.279,62.967){\line(0,1){.8923}}
\put(89.353,63.859){\line(0,1){1.1411}}
%\end
%\circle(187.5,65.25){38.275}
\put(206.637,65.25){\line(0,1){.8952}}
\put(206.617,66.145){\line(0,1){.8932}}
\put(206.554,67.038){\line(0,1){.8893}}
\multiput(206.449,67.928)(-.029199,.176689){5}{\line(0,1){.176689}}
\multiput(206.303,68.811)(-.031193,.145941){6}{\line(0,1){.145941}}
\multiput(206.116,69.687)(-.032559,.123705){7}{\line(0,1){.123705}}
\multiput(205.888,70.553)(-.033521,.106791){8}{\line(0,1){.106791}}
\multiput(205.62,71.407)(-.030784,.084085){10}{\line(0,1){.084085}}
\multiput(205.312,72.248)(-.03153,.075048){11}{\line(0,1){.075048}}
\multiput(204.965,73.073)(-.032089,.067367){12}{\line(0,1){.067367}}
\multiput(204.58,73.882)(-.0324971,.060731){13}{\line(0,1){.060731}}
\multiput(204.158,74.671)(-.0327807,.0549198){14}{\line(0,1){.0549198}}
\multiput(203.699,75.44)(-.0329595,.0497713){15}{\line(0,1){.0497713}}
\multiput(203.204,76.187)(-.0330483,.0451641){16}{\line(0,1){.0451641}}
\multiput(202.676,76.909)(-.0330586,.0410059){17}{\line(0,1){.0410059}}
\multiput(202.114,77.607)(-.0329994,.037225){18}{\line(0,1){.037225}}
\multiput(201.52,78.277)(-.032878,.0337648){19}{\line(0,1){.0337648}}
\multiput(200.895,78.918)(-.0344214,.03219){19}{\line(-1,0){.0344214}}
\multiput(200.241,79.53)(-.0378833,.0322415){18}{\line(-1,0){.0378833}}
\multiput(199.559,80.11)(-.0416647,.0322245){17}{\line(-1,0){.0416647}}
\multiput(198.851,80.658)(-.0458218,.0321303){16}{\line(-1,0){.0458218}}
\multiput(198.118,81.172)(-.0504262,.0319485){15}{\line(-1,0){.0504262}}
\multiput(197.361,81.651)(-.0555701,.0316658){14}{\line(-1,0){.0555701}}
\multiput(196.583,82.095)(-.0613744,.031265){13}{\line(-1,0){.0613744}}
\multiput(195.785,82.501)(-.074182,.033516){11}{\line(-1,0){.074182}}
\multiput(194.969,82.87)(-.083236,.033011){10}{\line(-1,0){.083236}}
\multiput(194.137,83.2)(-.094099,.032312){9}{\line(-1,0){.094099}}
\multiput(193.29,83.491)(-.107446,.03136){8}{\line(-1,0){.107446}}
\multiput(192.431,83.741)(-.124337,.030056){7}{\line(-1,0){.124337}}
\multiput(191.56,83.952)(-.146541,.028242){6}{\line(-1,0){.146541}}
\multiput(190.681,84.121)(-.22155,.03203){4}{\line(-1,0){.22155}}
\put(189.795,84.249){\line(-1,0){.8912}}
\put(188.904,84.336){\line(-1,0){.8943}}
\put(188.009,84.381){\line(-1,0){.8954}}
\put(187.114,84.384){\line(-1,0){.8946}}
\put(186.219,84.345){\line(-1,0){.8918}}
\multiput(185.327,84.264)(-.22175,-.03061){4}{\line(-1,0){.22175}}
\multiput(184.44,84.141)(-.176064,-.032758){5}{\line(-1,0){.176064}}
\multiput(183.56,83.978)(-.124528,-.029256){7}{\line(-1,0){.124528}}
\multiput(182.688,83.773)(-.107645,-.030668){8}{\line(-1,0){.107645}}
\multiput(181.827,83.527)(-.094305,-.031706){9}{\line(-1,0){.094305}}
\multiput(180.979,83.242)(-.083446,-.032474){10}{\line(-1,0){.083446}}
\multiput(180.144,82.917)(-.074396,-.033038){11}{\line(-1,0){.074396}}
\multiput(179.326,82.554)(-.066705,-.033442){12}{\line(-1,0){.066705}}
\multiput(178.525,82.153)(-.0600629,-.0337159){13}{\line(-1,0){.0600629}}
\multiput(177.744,81.714)(-.0506307,-.0316234){15}{\line(-1,0){.0506307}}
\multiput(176.985,81.24)(-.0460276,-.0318348){16}{\line(-1,0){.0460276}}
\multiput(176.249,80.731)(-.0418711,-.0319557){17}{\line(-1,0){.0418711}}
\multiput(175.537,80.187)(-.0380899,-.0319971){18}{\line(-1,0){.0380899}}
\multiput(174.851,79.611)(-.0346278,-.0319678){19}{\line(-1,0){.0346278}}
\multiput(174.193,79.004)(-.0330945,-.0335526){19}{\line(0,-1){.0335526}}
\multiput(173.564,78.367)(-.0332382,-.0370119){18}{\line(0,-1){.0370119}}
\multiput(172.966,77.7)(-.0333218,-.0407924){17}{\line(0,-1){.0407924}}
\multiput(172.4,77.007)(-.0333382,-.0449506){16}{\line(0,-1){.0449506}}
\multiput(171.866,76.288)(-.0332791,-.0495582){15}{\line(0,-1){.0495582}}
\multiput(171.367,75.544)(-.0331334,-.0547078){14}{\line(0,-1){.0547078}}
\multiput(170.903,74.778)(-.0328872,-.0605207){13}{\line(0,-1){.0605207}}
\multiput(170.476,73.992)(-.032522,-.067159){12}{\line(0,-1){.067159}}
\multiput(170.085,73.186)(-.032013,-.074844){11}{\line(0,-1){.074844}}
\multiput(169.733,72.362)(-.031324,-.083885){10}{\line(0,-1){.083885}}
\multiput(169.42,71.524)(-.030407,-.094732){9}{\line(0,-1){.094732}}
\multiput(169.146,70.671)(-.033355,-.123493){7}{\line(0,-1){.123493}}
\multiput(168.913,69.806)(-.032132,-.145738){6}{\line(0,-1){.145738}}
\multiput(168.72,68.932)(-.030335,-.176497){5}{\line(0,-1){.176497}}
\put(168.568,68.05){\line(0,-1){.8886}}
\put(168.458,67.161){\line(0,-1){.8928}}
\put(168.39,66.268){\line(0,-1){.895}}
\put(168.363,65.373){\line(0,-1){.8953}}
\put(168.378,64.478){\line(0,-1){.8936}}
\put(168.435,63.584){\line(0,-1){.89}}
\multiput(168.534,62.694)(.028062,-.176873){5}{\line(0,-1){.176873}}
\multiput(168.674,61.81)(.030254,-.146139){6}{\line(0,-1){.146139}}
\multiput(168.856,60.933)(.031763,-.123912){7}{\line(0,-1){.123912}}
\multiput(169.078,60.066)(.032834,-.107004){8}{\line(0,-1){.107004}}
\multiput(169.341,59.21)(.033603,-.093646){9}{\line(0,-1){.093646}}
\multiput(169.643,58.367)(.031047,-.075249){11}{\line(0,-1){.075249}}
\multiput(169.985,57.539)(.031655,-.067572){12}{\line(0,-1){.067572}}
\multiput(170.365,56.728)(.0321057,-.0609389){13}{\line(0,-1){.0609389}}
\multiput(170.782,55.936)(.0324267,-.0551296){14}{\line(0,-1){.0551296}}
\multiput(171.236,55.164)(.0326386,-.0499823){15}{\line(0,-1){.0499823}}
\multiput(171.725,54.414)(.0327571,-.0453758){16}{\line(0,-1){.0453758}}
\multiput(172.25,53.688)(.0327941,-.0412178){17}{\line(0,-1){.0412178}}
\multiput(172.807,52.988)(.0327592,-.0374365){18}{\line(0,-1){.0374365}}
\multiput(173.397,52.314)(.03266,-.0339757){19}{\line(0,-1){.0339757}}
\multiput(174.017,51.668)(.0342136,-.0324108){19}{\line(1,0){.0342136}}
\multiput(174.667,51.053)(.0376751,-.0324846){18}{\line(1,0){.0376751}}
\multiput(175.346,50.468)(.0414565,-.0324919){17}{\line(1,0){.0414565}}
\multiput(176.05,49.915)(.0456141,-.0324244){16}{\line(1,0){.0456141}}
\multiput(176.78,49.397)(.0502196,-.0322723){15}{\line(1,0){.0502196}}
\multiput(177.533,48.913)(.0553652,-.0320227){14}{\line(1,0){.0553652}}
\multiput(178.309,48.464)(.061172,-.0316593){13}{\line(1,0){.061172}}
\multiput(179.104,48.053)(.067801,-.03116){12}{\line(1,0){.067801}}
\multiput(179.917,47.679)(.083022,-.033545){10}{\line(1,0){.083022}}
\multiput(180.748,47.343)(.093889,-.032917){9}{\line(1,0){.093889}}
\multiput(181.593,47.047)(.107242,-.03205){8}{\line(1,0){.107242}}
\multiput(182.451,46.791)(.124141,-.030856){7}{\line(1,0){.124141}}
\multiput(183.319,46.575)(.146356,-.029184){6}{\line(1,0){.146356}}
\multiput(184.198,46.4)(.22134,-.03346){4}{\line(1,0){.22134}}
\put(185.083,46.266){\line(1,0){.8907}}
\put(185.974,46.173){\line(1,0){.894}}
\put(186.868,46.123){\line(1,0){.8954}}
\put(187.763,46.114){\line(1,0){.8948}}
\put(188.658,46.148){\line(1,0){.8923}}
\put(189.55,46.223){\line(1,0){.8878}}
\multiput(190.438,46.339)(.176271,.031625){5}{\line(1,0){.176271}}
\multiput(191.319,46.497)(.145499,.033196){6}{\line(1,0){.145499}}
\multiput(192.192,46.697)(.10784,.029974){8}{\line(1,0){.10784}}
\multiput(193.055,46.936)(.094507,.031099){9}{\line(1,0){.094507}}
\multiput(193.906,47.216)(.083654,.031937){10}{\line(1,0){.083654}}
\multiput(194.742,47.536)(.074607,.032559){11}{\line(1,0){.074607}}
\multiput(195.563,47.894)(.066919,.033012){12}{\line(1,0){.066919}}
\multiput(196.366,48.29)(.0602786,.0333288){13}{\line(1,0){.0602786}}
\multiput(197.149,48.723)(.0544641,.0335325){14}{\line(1,0){.0544641}}
\multiput(197.912,49.193)(.0493135,.0336405){15}{\line(1,0){.0493135}}
\multiput(198.652,49.697)(.0447056,.033666){16}{\line(1,0){.0447056}}
\multiput(199.367,50.236)(.0405477,.0336191){17}{\line(1,0){.0405477}}
\multiput(200.056,50.808)(.0367679,.0335079){18}{\line(1,0){.0367679}}
\multiput(200.718,51.411)(.0333097,.033339){19}{\line(0,1){.033339}}
\multiput(201.351,52.044)(.0334757,.0367973){18}{\line(0,1){.0367973}}
\multiput(201.953,52.706)(.0335835,.0405772){17}{\line(0,1){.0405772}}
\multiput(202.524,53.396)(.0336268,.0447351){16}{\line(0,1){.0447351}}
\multiput(203.062,54.112)(.0335972,.049343){15}{\line(0,1){.049343}}
\multiput(203.566,54.852)(.0334847,.0544935){14}{\line(0,1){.0544935}}
\multiput(204.035,55.615)(.0332759,.0603078){13}{\line(0,1){.0603078}}
\multiput(204.468,56.399)(.032953,.066948){12}{\line(0,1){.066948}}
\multiput(204.863,57.202)(.032493,.074636){11}{\line(0,1){.074636}}
\multiput(205.221,58.023)(.031863,.083682){10}{\line(0,1){.083682}}
\multiput(205.539,58.86)(.031016,.094534){9}{\line(0,1){.094534}}
\multiput(205.818,59.711)(.02988,.107866){8}{\line(0,1){.107866}}
\multiput(206.057,60.574)(.033069,.145528){6}{\line(0,1){.145528}}
\multiput(206.256,61.447)(.03147,.176299){5}{\line(0,1){.176299}}
\put(206.413,62.329){\line(0,1){.8879}}
\put(206.529,63.217){\line(0,1){.8923}}
\put(206.603,64.109){\line(0,1){1.1411}}
%\end
%\vector(70.5,65)(82.75,79.75)
\put(82.75,79.75){\vector(3,4){.07}}\multiput(70.5,65)(.0336538462,.040521978){364}{\line(0,1){.040521978}}
%\end
%\vector(187.75,65.25)(200,80)
\put(200,80){\vector(3,4){.07}}\multiput(187.75,65.25)(.0336538462,.040521978){364}{\line(0,1){.040521978}}
%\end
\put(75.5,75.25){\makebox(0,0)[cc]{$R$}}
\put(192.75,75.5){\makebox(0,0)[cc]{$R$}}
\linethickness{0.1pt}
%\emline(52.25,58.75)(61,81.5)
\multiput(52.25,58.75)(.0336538462,.0875){260}{\line(0,1){.0875}}
%\end
%\emline(169.5,58.75)(178.25,81.5)
\multiput(169.5,58.75)(.0336538462,.0875){260}{\line(0,1){.0875}}
%\end
%\emline(54.75,54.25)(66.25,83.75)
\multiput(54.75,54.25)(.0337243402,.0865102639){341}{\line(0,1){.0865102639}}
%\end
%\emline(172,54.25)(183.5,83.75)
\multiput(172,54.25)(.0337243402,.0865102639){341}{\line(0,1){.0865102639}}
%\end
%\emline(58.25,51)(71,84)
\multiput(58.25,51)(.0337301587,.0873015873){378}{\line(0,1){.0873015873}}
%\end
%\emline(175.5,51)(188.25,84)
\multiput(175.5,51)(.0337301587,.0873015873){378}{\line(0,1){.0873015873}}
%\end
%\emline(61.75,48)(75,83.25)
\multiput(61.75,48)(.0337150127,.0896946565){393}{\line(0,1){.0896946565}}
%\end
%\emline(179,48)(192.25,83.25)
\multiput(179,48)(.0337150127,.0896946565){393}{\line(0,1){.0896946565}}
%\end
%\emline(65.75,46.5)(79.25,82)
\multiput(65.75,46.5)(.0336658354,.0885286783){401}{\line(0,1){.0885286783}}
%\end
%\emline(183,46.5)(196.5,82)
\multiput(183,46.5)(.0336658354,.0885286783){401}{\line(0,1){.0885286783}}
%\end
%\emline(70,46.5)(83.25,78.75)
\multiput(70,46.5)(.0337150127,.0820610687){393}{\line(0,1){.0820610687}}
%\end
%\emline(187.25,46.5)(200.5,78.75)
\multiput(187.25,46.5)(.0337150127,.0820610687){393}{\line(0,1){.0820610687}}
%\end
%\emline(75.25,47)(86.5,74)
\multiput(75.25,47)(.0336826347,.0808383234){334}{\line(0,1){.0808383234}}
%\end
%\emline(192.5,47)(203.75,74)
\multiput(192.5,47)(.0336826347,.0808383234){334}{\line(0,1){.0808383234}}
%\end
%\emline(81.5,49.75)(88.75,67)
\multiput(81.5,49.75)(.03372093,.080232558){215}{\line(0,1){.080232558}}
%\end
%\emline(198.75,49.75)(206,67)
\multiput(198.75,49.75)(.03372093,.080232558){215}{\line(0,1){.080232558}}
%\end
%\emline(27,72.25)(20.75,57.75)
\multiput(27,72.25)(-.033602151,-.077956989){186}{\line(0,-1){.077956989}}
%\end
%\emline(236.25,73.5)(230,59)
\multiput(236.25,73.5)(-.033602151,-.077956989){186}{\line(0,-1){.077956989}}
%\end
%\emline(31.75,71.25)(26.75,59)
\multiput(31.75,71.25)(-.033557047,-.082214765){149}{\line(0,-1){.082214765}}
%\end
%\emline(230.25,71.75)(225.25,59.5)
\multiput(230.25,71.75)(-.033557047,-.082214765){149}{\line(0,-1){.082214765}}
%\end
%\emline(36.5,70.75)(32.25,60)
\multiput(36.5,70.75)(-.033730159,-.08531746){126}{\line(0,-1){.08531746}}
%\end
%\emline(225,70.75)(220.75,60)
\multiput(225,70.75)(-.033730159,-.08531746){126}{\line(0,-1){.08531746}}
%\end
%\emline(40.5,70)(37,60.75)
\multiput(40.5,70)(-.033653846,-.088942308){104}{\line(0,-1){.088942308}}
%\end
%\emline(220.5,70)(217,60.75)
\multiput(220.5,70)(-.033653846,-.088942308){104}{\line(0,-1){.088942308}}
%\end
%\emline(45.5,69.25)(42.5,61.25)
\multiput(45.5,69.25)(-.03370787,-.08988764){89}{\line(0,-1){.08988764}}
%\end
%\emline(215.5,69.25)(212.5,61.25)
\multiput(215.5,69.25)(-.03370787,-.08988764){89}{\line(0,-1){.08988764}}
%\end
%\emline(50.5,68.75)(48.25,62.75)
\multiput(50.5,68.75)(-.03358209,-.08955224){67}{\line(0,-1){.08955224}}
%\end
%\emline(210.75,68.75)(208.5,62.75)
\multiput(210.75,68.75)(-.03358209,-.08955224){67}{\line(0,-1){.08955224}}
%\end
\put(101,76.25){\makebox(0,0)[cc]{$\cD$}}
\put(152.5,76.75){\makebox(0,0)[cc]{$\cD_1$}}
\thicklines
\linethickness{0.5mm}
\put(80,73){\line(0,-1){14}}
\put(178.5,73){\line(0,-1){14}}
\put(59,69.0){\line(0,-1){6.75}}
\put(199.5,69.0){\line(0,-1){6.75}}
\put(24.25,65.55){\line(1,0){6.25}}
\put(234.25,65.55){\line(-1,0){6.25}}
\put(36.5,65.55){\line(1,0){6.5}}
\put(222,65.55){\line(-1,0){6.5}}
\put(50.5,66.75){\line(0,-1){2.5}}
\put(208,66.75){\line(0,-1){2.5}}
\end{picture}
\caption{\footnotesize
The domain $\cD$ of analyticity of the function $\p_3(x,\cdot)$
(fig.~a), the domain $\cD_1$ of analyticity of the function $\p_1(x,\cdot)$ (fig.~b),
and the slits for the good Boussinesq equation
}
\label{Figbd}
\end{figure}
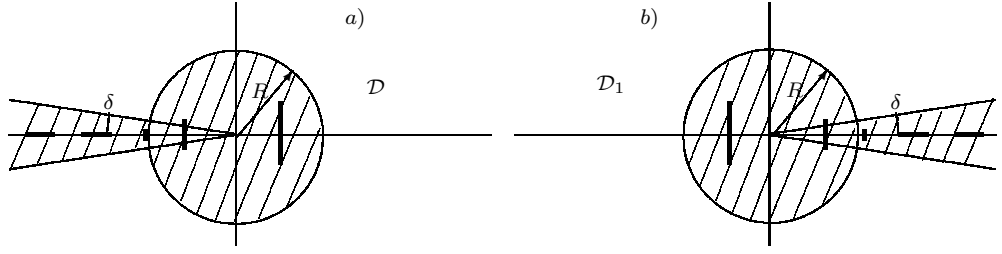

\subsection{Transformation to a second order equation}
An important problem is to prove that the high energy ramifications are real.
In order to solve this problem McKean (referring to J.Moser)
reduces the third-order equation \er{1b} to a second-order equation with
an energy-dependent potential. Now we describe this transformation.

Each function $\vk_3$ and $\p_3(x,\cdot),x\in\R$, is analytic in the domain
$$
\cD=\{\zeta\in\C:|\zeta|>R,|\arg\zeta|<\pi-\d\}
$$
(see Fig.~\ref{Figbd}~a)
for any $\d>0$ small enough and for some
$R>0$ large enough. Moreover, if $\zeta\to\iy$ in $\cD$, then
$$
\vk_3(\zeta)=e^{\zeta}(1+O(|\zeta|^{-1})),\qqq
\p_3(x,\zeta)=e^{x\zeta}(1+O(|\zeta|^{-1}))
$$
uniformly in $x\in[0,1]$.
Therefore, $|\vk_3|$ and $|\p_3(x,\cdot)|,x\in\R$,
are increasing as $|\zeta|\to\iy$ in $\cD$.
Using this result we take $R>0$ so large that the function
$\p_3(x,\zeta)$ does not vanish in $\R\ts\cD$.

Let $\zeta\in\cD$.
If we take any solution $y$ of equation \er{1b}, then the function
\[
\lb{deff}
f=\p_3^{3\/2}\Big({y\/\p_3}\Big)'
\]
satisfies the equation
\[
\lb{2oeq}
-f''+\cV f=0,
\]
where the energy-dependent potential
$\cV(x,\zeta)$ has the form
\[
\lb{da1}
\cV=-2p-{3\/4}\Big(2\big({\p_3'\/\p_3}\big)'+\big({\p_3'\/\p_3}\big)^2\Big),
\]
and satisfies $\cV(\cdot,\zeta)\in L^1(\T)$.
Each function $\cV(x,\cdot),x\in\R$, is analytic in the domain
$\cD$, real on $\R\cap\cD$, and satisfies
\[
\lb{asa}
\cV(x,\zeta)=-\l-p(x)+O(\zeta^{-{1\/3}}),\qqq\l={3\/4}\zeta^{2\/3},
\]
as $|\zeta|\to\iy, \zeta\in\cD$, uniformly on $x\in\T$.
Then equation \er{2oeq} has the form \er{2oequd}, where
\[
\lb{VBe}
 V(x,\l)=\cV(x,\zeta)+\l.
\]
For each $x\in\R$ the function $V(x,\cdot)$ is analytic in the domain $\mD$
given by
\[
\lb{mDBe}
\mD=\Big\{\l\in\C:|\l|>{3\/4}R^{2\/3},|\arg\l|<{2\/3}(\pi-\d)\Big\}.
\]
The asymptotics \er{asa} shows that
$V(x,\l)=-p(x)+O(\l^{-{1\/2}})$
as $|\l|\to\iy$ in $\mD$, uniformly in $x\in[0,1]$.

\subsection{Results for the Boussinesq equation}
Introduce the fundamental solutions $\phi_1(x,\zeta)$, $\phi_2(x,\zeta)$,
$(x,\zeta)\in\R\ts\cD$
of the equation \er{2oeq}, satisfying the conditions $\phi_1(0,\zeta)=\phi_2'(0,\zeta)=1$,
$\phi_1'(0,\zeta)=\phi_2(0,\zeta)=0$. Introduce the fundamental matrix
$\Phi=(\phi_j^{(k-1)})_{j,k=1}^2$.
The matrix $\Phi(1,\zeta),\zeta\in\cD$, is the monodromy matrix.
It is analytic in $\cD$.
It has two eigenvalues $\t_1(\zeta),\t_2(\zeta)$, they are the multipliers.
The multipliers satisfy the identity $\t_1\t_2=1$.
The discriminant $(\t_1-\t_2)^2=2(\phi_1(1,\cdot)+\phi_2'(1,\cdot))$ is an analytic function in $\cD$.
The zeros of this function are the eigenvalues
of the 2-periodic problem for equation \er{2oeq}.
For each multiplier
there are the Floquet solution $f_j(x,\zeta),j=1,2$,
satisfying the conditions
$
f_j(x+1,\zeta)=\t_j(\zeta) f_j(x,\zeta).
$

In our next article, we will show that the ramifications of the multiplier surface $\mR$
coincide with the eigenvalues
of the 2-periodic problem for equation \er{2oeq},
and the three-point Dirichlet eigenvalues coincide with the Dirichlet eigenvalues
for equation \er{2oeq}.
Here we briefly describe the corresponding arguments.

Let $y=\p_1$ be the Floquet solution of equation \er{1b}.
Then $f_1=\p_3^{3\/2}({\p_1\/\p_3})'$
is the Floquet solution of equation \er{2oeq} satisfying
$f_1(x+1)=\vk_3^{1\/2}\vk_1f_1(x)$.
Similarly, $f_2=\p_3^{3\/2}({\p_2\/\p_3})'$
is the Floquet solution of equation \er{2oeq} satisfying
$f_2(x+1)=\vk_3^{1\/2}\vk_2f_2(x)$.
Then $\t_1=\vk_3^{1\/2}\vk_1$
and $
\t_2=\vk_3^{1\/2}\vk_2
$
are multipliers for equation \er{2oeq}.

Let $\zeta\in\cD$ be a ramification of the surface $\mR$.
Recall that in this case at least two function $\vk_j,j=1,2,3$, take the same value.
The identity $\vk_1\vk_2\vk_3=1$ and the asymptotics
$\vk_3(\zeta)=e^{\zeta}(1+O(|\zeta|^{-1}))$ show that
$\vk_1(\zeta)=\vk_2(\zeta)$, which yields
$\t_1(\zeta)=\t_2(\zeta)$.
Therefore, $\zeta$ is
an eigenvalue of the 2-periodic problem for equation \er{2oeq}.

Furthermore, if $\zeta\in\cD$ is an eigenvalue of the three-point Dirichlet problem
for equation \er{1b}, then it is a pole of the Floquet solution $\p_1$ or $\p_2$
of equation \er{1b}, therefore, it is a pole of the Floquet solution $f_1$ or $f_2$
of equation \er{2oeq}. Then it is an eigenvalue of the Dirichlet problem
for equation \er{2oeq}.

In the following corollary of the previous theorems (see the proof in Section~6)
we extend McKean's result that the ramifications are real
from the class $p,q\in C^\iy(\T)$ onto a wider class of coefficients $p',q\in L^1(\T)$.

\begin{corollary}
\lb{Corbe}
Let $p',q\in L^1(\T)$. Then

i) The ramifications and the eigenvalues of the three-point problem
in the half-plane $\cZ_a=\{\zeta\in\C:\Re\zeta>a\}$ are real for some $a>0$ large enough.
There are exactly two (counting with multiplicity) ramifications $r_{n}^\pm$
and exactly one simple three-point eigenvalue $\zeta_n$
in each interval $(\a_n^-,\a_n^+)$ inside this half-plane,
where $n\in\N,\a_n^\pm=({\pi (2n\pm1)\/\sqrt3})^3$.
There are no other ramifications and  three-point eigenvalues in the half-plane $\cZ_a$.

ii) The eigenvalues $\zeta_n$ satisfy
\[
\lb{as3p}
\zeta_n\in[r_n^-,r_n^+]\ss\R,
\]
for all $n\in\N$ large enough.

iii) The ramifications $r_n^\pm$ satisfy
\[
\lb{asram}
r_{n}^\pm=\Big({2\pi n\/\sqrt3}\Big)^3-{4\pi np_0\/\sqrt3}+o(n),
\]
as $n\to+\iy$, where $p_0=\int_0^1p(x)dx$.

\end{corollary}

\no {\bf Remark.} 1) Similarly the negative $\zeta$ may be considered.
The Floquet solution $\p_1$ is analytic in the domain
$\cD_1=\{\zeta\in\C:|\zeta|>R,|\arg\zeta|>\d\}$, see Fig.~\ref{Figbd}~b.
 If we use this function instead of $\p_3$
in the previous construction, then we obtain the relations
similar to \er{asram} and \er{as3p} for $n\to-\iy$.

2) The multiplier Riemann surface and the ramifications for the self-adjoint
third order operator associated with the bad Boussinesq equation was the subject
of our papers \cite{BK14}, \cite{BK15}.
The multiplier surface and the ramifications for the good Boussinesq
are the subjects of our next paper.

3) The relations \er{as3p} are proved by McKean \cite{McK81} for the smooth
coefficients $p,q$. Our prove is simpler and extends these relations
onto the larger class of the coefficients $p',q\in L^1(\T)$.

4) Assuming a higher smoothness of the coefficients,
we can improve the asymptotics \er{asram} in order to determine a trace formula.
This is the subject of our next paper.

5) The previous results may be extended from the class $p',q\in L^1(\T)$
onto the class $p,q\in L^1(\T)$. The transformation \er{deff} in this case
leads to the potential $\cV$ that is the distribution
with respect to $x$.
Then we have to consider equation \er{2oequd} where the potential $V$ is a distribution.
We think that our results hold for this case.
The corresponding energy-independent potentials were considered
by Korotyaev \cite{K03}.

6) The sharp asymptotics of the ramifications for the bad Boussinesq equation
is determined in \cite{BK15}.

\section{The Lyapunov function and the spectra}
\setcounter{equation}{0}

\subsection{The fundamental solutions}
Introduce  the fundamental solutions $\vt(x,\l)$, $\vp(x,\l)$, $(x,\l)\in\R\ts\mD$,
of equation \er{2oequd}
satisfying the initial conditions $\vt(0,\l)=\vp'(0,\l)=1,\vt'(0,\l)=\vp(0,\l)=0$.

The fundamental solutions $\vt(x,\l),\vp(x,\l)$ of the unperturbed equation $-y''=\l y$
have the form
$$
\vt_o(x,\l)=\cos zx,\qqq \vp_o(x,\l)={\sin zx\/z},\qqq z=\sqrt\l,
$$
here and below  $\sqrt1=1$.
Each function $ \vt_o(x,\cdot)$, $ \vp_o(x,\cdot)$, $x\in\R$, is entire.

Each solution $y(x,\l),(x,\l)\in\R_+\ts\mD$, of equation \er{2oequd} satisfies the following integral
equation
$$
y(x,\l)=y(0,\l)\vt_o(x,\l)+y'(0,\l)\vp_o(x,\l)+\int_0^x\vp_o(x-s,\l) V(s,\l)y(s,\l)ds,
\qq x\in\R.
$$
The standard iterations give
\[
\lb{deffs}
\begin{aligned}
\vt(x,\l)=\sum_{n=0}^\iy \vt_n(x,\l),\qq
\vt_{n}(x,\l)=\int_0^x\vp_o(x-s,\l) V(s,\l)\vt_{n-1}(s,\l)ds,
\\
\vp(x,\l)=\sum_{n=0}^\iy \vp_n(x,\l),\qq
\vp_{n}(x,\l)=\int_0^x\vp_o(x-s,\l) V(s,\l)\vp_{n-1}(s,\l)ds,
\end{aligned}
\]
for each $(n,x,\l)\in\N\ts\R_+\ts\mD$.

\begin{lemma}
\lb{Lmanfm}
Each function
$\vt(x,\cdot)$, $\vp(x,\cdot)$, $\vt'(x,\cdot)$, $\vp'(x,\cdot),x\in\R$,
is analytic in $\mD$.
Moreover,
\[
\lb{estfs}
\begin{aligned}
&\sup\Big\{
\Big|\vt(x,\l)-\sum_{n=0}^N \vt_n(x,\l)\Big|,
|z|_1\Big|\vp(x,\l)-\sum_{n=0}^N \vp_n(x,\l)\Big|,
\\
&\Big|\vp'(x,\l)-\sum_{n=0}^N \vp_n'(x,\l)\Big|,
{1\/|z|_1}\Big|\vt'(x,\l)-\sum_{n=0}^N \vt_n'(x,\l)\Big|
\Big\}
\le {\| V(\cdot,\l)\|^{N+1}e^{x|\Im z|+{\| V(\cdot,\l)\|\/|z|_1}}\/|z|_1^{N+1}},
\end{aligned}
\]
for all $N\ge 0,(x,\l)\in\R_+\ts\mD$, where $|z|_1=\max\{1,|z|\}$.
\end{lemma}

\no {\bf Proof.} The standard arguments, see, e.g., \cite[Ch~1,Thms~1 and 3]{PT87},
give
$$
\vt_{n}(x,\l)={1\/z^n}\int\limits_{0<x_1<...<x_{n}<x_{n+1}=x}
\prod_{k=1}^{n}\sin z(x_{k+1}-x_k) V(x_k,\l)\cos zx_1dx_1...dx_n,
$$
which yields
$$
\begin{aligned}
|\vt_{n}(x,\l)|\le{1\/|z|_1^n}\int\limits_{0<x_1<...<x_{n}<x_{n+1}=x}
\prod_{k=1}^{n}e^{|\Im z|(x_{k+1}-x_k)}| V(x_k,\l)|e^{|\Im z|x_1}dx_1...dx_n
\\
\le {e^{|\Im z|x}\/|z|_1^n}\int\limits_{0<x_1<...<x_{n}<x_{n+1}=x}
\prod_{k=1}^{n}| V(x_k,\l)|dx_1...dx_n=
{e^{|\Im z|x}\/n!|z|_1^n}\Big(\int_0^x| V(x,\l)|dx\Big)^n.
\end{aligned}
$$
This estimate and the similar estimate for
$|\vp_n(x,\l)|,|\vt_n'(x,\l)|,|\vp_n'(x,\l)|$
imply
$$
\max\{|\vt_n(x,\l)|,|z|_1|\vp_n(x,\l)|,
|z|_1^{-1}|\vt_n'(x,\l)|,|z|_1|\vp_n'(x,\l)|\}
\le{\| V(\cdot,\l)\|^n\/n!|z|_1^n}e^{x|\Im z|},
$$
for all $n\ge 0,(x,\l)\in\R_+\ts\mD.$
These estimates show that
the series converge uniformly on any compact in $\mD$ and the sums are
analytic functions in $\mD$ and satisfy
$$
\max\{|\vt(x,\l)|,|\vp(x,\l)|\}\le e^{x|\Im z|+{\| V(\cdot,\l)\|\/|z|_1}},\qq
\qq \forall\ \ (x,\l)\in\R_+\ts\mD.
$$
Summing the majorants we obtain the estimates \er{estfs}.
\BBox

\subsection{Asymptotics of the Lyapunov function}

Introduce the Lyapunov function by
\[
\lb{fl2ovv}
\D(\l)={1\/2}\big(\vt(1,\l)+\vp'(1,\l)\big),\qqq \l\in \mD.
\]
The function $\D$ is analytic in $\mD$.

In the following Lemma we prove estimates for the solution $\vp(1,\l)$ and the Lyapunov function
$\D(\l)$. Introduce the functions
\[
\lb{defDj}
\D_j(\l)={1\/2}\big(\vt_j(1,\l)+\vp_j'(1,\l)\big),\qq j\in\N,\qq \l\in\mD.
\]

\begin{lemma}
i) The functions $\D_j,j=1,2$, satisfy
\[
\lb{idD1}
\D_1(\l)={\sin z\/2z}\wh V_o(\l),
\]
\[
\lb{idD2}
\begin{aligned}
\D_2(\l)={1\/4z^2}\lt(\cos z\Big(\int_0^1ds\int_0^s\cos2 z(s-t) W(s,t,\l)dt
-{\wh V_o^2(\l)\/2}\Big)
\\
+\sin z\int_0^1ds\int_0^s\sin 2z(s-t) W(s,t,\l)dt\rt),
\end{aligned}
\]
for all $\l\in\mD$, where $\wh V_o(\l)=\int_0^1 V(s,\l)ds$ and $W(s,t,\l)=V(s,\l)V(t,\l)$.

ii) The following estimates hold true:
\[
\lb{estfir}
\Big|\vp(1,\l)-{\sin z\/z}\Big|\le
{e_1(\l)\/|z|_1},
\qq
\]
\[
\lb{estLfr}
\big|\D(\l)-\cos z\big|
\le e_1(\l),
\]
\[
\lb{asLfm}
\big|\D(\l)-\cos z-\D_1(\l)\big|
\le e_2(\l),
\]
\[
\lb{asLfsp}
\big|\D(\l)-\cos z-\D_1(\l)-\D_2(\l)\big|\le e_3(\l),
\]
for all $\l\in\mD$, where
$$
e_j(\l)={\| V(\cdot,\l)\|^{j}\/|z|_1^{j}}e^{|\Im z|+{\| V(\cdot,\l)\|\/|z|_1}},\qq
j\ge 0.
$$
\end{lemma}

\no {\bf Proof.}
i) The definitions \er{deffs} imply
$$
\vt_1(1)={1\/z}\int_0^1 V(s)\sin z(1-s)\cos zsds,
\qq
\vp_1'(1)={1\/z}\int_0^1V(s)\cos z(1-s)\sin zsds,
$$
here and below in this proof $\vt(x)=\vt(x,\l),V(x)=V(x,\l),W(s,t)=W(s,t,\l),...$
Substituting these identities into the definition \er{defDj} we obtain \er{idD1}.
%\[
%\lb{trmm1}
%\vt_1(1)+\vp_1'(1)={\sin z\/z}\wh V_o,
%\]
Moreover,
$$
\begin{aligned}
\vt_2(1)={1\/z^2}\int_0^1ds\int_0^sW(s,t)\sin z(1-s)\sin z(s-t)\cos ztdt,
\\
\vp_2'(1)={1\/z^2}\int_0^1ds\int_0^sW(s,t)\cos z(1-s)\sin z(s-t)\sin ztdt,
\end{aligned}
$$
therefore,
$$
\begin{aligned}
&2\D_1=\vt_2(1)+\vp_2'(1)
={1\/z^2}\int_0^1ds\int_0^sW(s,t)\sin z(1-s+t)\sin z(s-t)dt
\\
&={1\/2z^2}\Big(\sin z\int_0^1ds\int_0^s\sin 2z(s-t)W(s,t)dt
-\cos z\int_0^1ds\int_0^s\big(1-\cos2 z(s-t)\big)W(s,t)dt\Big),
%\\
%&={1\/2z^2}\lt(\cos z\Big(\int_0^1ds\int_0^s\cos2 z(s-t)W(s,t)dt
%-{\wh V_o^2\/2}\Big)
%+\sin z\int_0^1ds\int_0^s\sin 2z(s-t) W(s,t)dt\rt),
\end{aligned}
$$
which yields \er{idD2}.

ii) Let $\l\in\mD$. The estimates \er{estfs} give \er{estfir} and
$$
\max\big\{|\vt(1,\l)-\cos z|,
|\vp'(1,\l)-\cos z|\big\}
\le e_1(\l),
$$
$$
\max\Big\{\Big|\vt(1,\l)-\cos z-\vt_1(1,\l)\Big|,
\Big|\vp'(1,\l)-\cos z-\vp_1'(1,\l)\Big|\Big\}
\le e_2(\l),
$$
$$
\max\Big\{\Big|\vt(1,\l)-\cos z-\vt_1(1,\l)
-\vt_2(1,\l)\Big|,
\Big|\vp'(1,\l)-\cos z-\vp_1'(1,\l)
-\vp_2'(1,\l)\Big|\Big\}
\le e_3(\l).
$$
These estimates together with the definitions \er{defDj} yield \er{estLfr}--\er{asLfsp}.
\BBox

\subsection{The resolvent and the spectrum of the quasi-periodic problem}
The following Lemma shows
that the resolvents $(H(k,\l)-\l)^{-1}$
and  $(T(\l)-\l)^{-1}$
are similar to the corresponding resolvents
for the case of the potential independent of $\l$.

\begin{lemma}
\lb{LmResol}

i) Let $k\in[0,2\pi)$.
The set $\rho(H(k,\cdot))$ of all regular points of the operator $H(k,\l)$ satisfies
\[
\lb{rsH}
\rho(H(k,\cdot))=\{\l\in\mD:\D(\l)\ne\cos k\}.
\]
Let, in addition, $\l\in\rho(H(k,\cdot))$.
Then the resolvent $\cR_H(k,\l)=(H(k,\l)-\l)^{-1}$ is a bounded operator and
has the form
\[
\lb{sngp}
(\cR_H(k,\l)f)(x)=\int_0^1 R_H(x,s;k,\l)f(s)ds,\qq \forall\qq x\in\R,
\]
where
$$
R_H(x,s;k,\l)={\vp(1,\l)\/2(\cos k-\D(\l))} \ca
\p_-(s,\l)\vt(x,\l) +m_+(\l)\p_-(s,\l)\vp(x,\l)
,& s<x\\
\p_+(x,\l)\vt(s,\l) +m_-(\l)\p_+(x,\l)\vp(s,\l)
,&s>x \ac,
$$
$$
\begin{aligned}
\p_\pm(x,\l)=\vt(x,\l)+m_\pm(\l)\vp(x,\l), \qq
m_\pm(\l)={1\/\vp(1,\l)}\Big({\vp'(1,\l)-\vt(1,\l)\/2}\pm i\sin k\Big).
\end{aligned}
$$
The spectrum
$\s(H(k,\cdot))$
of the function $H(k,\l)$
is discrete
and coincides with the set
\[
\lb{splf}
\s(H(k,\cdot))=
\{\l\in\mD:\D(\l)=\cos k\}.
\]
The spectrum
of the 2-periodic problem has the form
\[
\lb{speck2p}
\s(H(0,\cdot))\cup\s(H(\pi,\cdot))
=\{\l\in\mD:\D(\l)=\pm1\}.
\]

ii) The set $\rho(T)$ of all regular points of the operator $T(\l)$ satisfies
\[
\lb{rsT}
\rho(T)=\{\l\in\mD:\vp(1,\l)\ne 0\}.
\]
Let $\l\in\rho(T)$.
Then the resolvent $\cR_T(\l)=(T(\l)-\l)^{-1}$ is a bounded operator and
has the form
\[
\lb{sngd}
(\cR_T(\l)f)(x)=\int_0^1 R_T(x,s;\l)f(s)ds,\qq \forall\qq x\in\R,
\]
where
$$
R_T(x,s;\l)={1\/\vp(1,\l)}
\ca\vp(s,\l)\big(\vt(1,\l)\vp(x,\l)-\vt(x,\l)\vp(1,\l)\big),& s<x\\
\vp(x,\l)\big(\vt(1,\l)\vp(s,\l)-\vt(s,\l)\vp(1,\l)\big),&s>x \ac.
$$
The spectrum $\s(T)$ of the operator $T(\l)$ is discrete and coincides with the set
\[
\lb{specd}
\s(T)=\{\l\in\mD:\vp(1,\l)=0\}.
\]

iii) The spectrum $\s(\cN)$ of the operator $\cN(\l)$ is discrete and coincides with the set
\[
\lb{specn}
\s(T)=\{\l\in\mD:\vt'(1,\l)=0\}.
\]
\end{lemma}

\no {\bf Proof.}
i) Direct calculations show that $R_H(x,s;k,\l)$
satisfies the standard properties of Green's functions
for equation \er{2oequd} and the conditions \er{fbc}.
This yields the identity \er{sngp}.
The  identity \er{sngp} shows that the resolvent is a bounded operator
for all $\l\in\mD$ such that $\D(\l)\ne\cos k$,
therefore the set of the regular points has the form \er{rsH} and then
the spectrum satisfies \er{splf}.
Let $\l\in\mD$ be a zero of the function $\D(\l)-\cos k$.
Then the function $\vp(1,\l)\vt(x,\l)-(\vt(1,\l)-e^{ik})\vp(x,\l)$
satisfies equation \er{2oequd} and the conditions \er{fbc}.
Therefore, $\l$ is an eigenvalue. Thus, the spectrum is pure discrete.
The identity \er{splf} yields \er{speck2p}.

ii)  The function $R_T(x,s,\l)$ is the Green function
for the problem \er{2oequd}, \er{dbc}.
This yields the identity \er{sngd}.
This  identity gives that the resolvent is a bounded operator
for all $\l\in\mD:\vp(1,\l)\ne 0$,
therefore the set of the regular points is given by \er{rsT}
and the spectrum satisfies \er{specd}.
Let $\l\in\mD$ be a zero of the function $\vp(1,\l)$.
Then the function $\vp(x,\l)$ is an eigenfunction of
the problem \er{2oequd}, \er{dbc} with the eigenvalue
$\l$. This yields that the spectrum is pure discrete.

iii) The proof is similar.
\BBox

\medskip

%\subsection{Lyapunov function}

The maximum number of linearly independent eigenvectors associated with an eigenvalue,
is referred to as the eigenvalue's geometric multiplicity.
The non-trivial solutions $y_1,y_2,...,y_{m-1}$ of the
equations
$$
\sum_{j=0}^n{1\/j!}(H(k,\l)-\l)^{(j)}|_{\l=\l_o}y_{n-j}=0,\qq n=1,2,...,m-1,
$$
are called
the {\it adjoined vectors} to the eigenvector $y_o$ and the number $m$ is called
the {\it algebraic multiplicity} of the eigenvalue $\l_o$.

The identity \er{splf} shows that the spectrum $\s(H(k,\cdot))$ of the quasiperiodic
problem consists of eigenvalues that are zeros of the function
$\D(\l)-\cos k$ analytic in $\mD$. Similarly, the identity
\er{specd} yields that the spectrum $\s(H_{d})$ of the Dirichlet
problem consists of eigenvalues that are zeros of the function
$\vp(1,\cdot)$ analytic in $\mD$. The multiplicity of the zero
is the algebraic multiplicity of the corresponding eigenvalue.
The algebraic and geometric multiplicity of the eigenvalue
can be different from each other.

We are ready to prove our results about the direct integral decomposition and the spectrum
of the operator $H(\l)$.

\medskip

\no {\bf Proof of Proposition~\ref{ThSpec}.}
i) The proof of the identity \er{deH} is standard, see \cite[Ch~XIII.16]{RS78}.

ii) The statement is proved in Lemma~\ref{LmResol}.

iii) The identity \er{splf} and the analyticity of the function $\D$
on the domain $\mD$ yield the statement.

iv) The decomposition \er{deH} and the statement iii) yield \er{specH}.
\BBox

\section{Conditions when the spectra are real.}
\setcounter{equation}{0}

\subsection{Local conditions}

Introduce the function
\[
\lb{defb}
\eta(x,\l)= Q(x,\l)-\nu,
\]
recall that $\l=\mu+i\nu$ and $ Q=\Im V$.
For each $x\in[0,1]$ the function $\eta(x,\cdot)$ is harmonic in $\mD$.
Below we need the following auxiliary result.

\begin{lemma}
\lb{LmVis0}
Let $\l_o=\mu_o+i\nu_o\in\gS$.
Then there exists (maybe not unique) point
$x_o=x_o(\l_o)\in(0,1)$ such that $\eta(x_o,\l_o)=0$.
Moreover, in this case
\[
\lb{qest}
|\nu_o|\le\sup_{x\in[0,1]}|Q(x,\l_o)|.
\]
\end{lemma}

\no {\bf Proof.}
We consider the spectrum $\s(H(k,\cdot))$ of the quasi-periodic problem.
The proofs for the Dirichlet spectrum $\s(T)$
and for the Neumann spectrum $\s(\cN)$ are similar.
Let $\l_o\in\s(H(k,\cdot))$.
The corresponding eigenfunction $y$ satisfies
$$
\begin{aligned}
0=\int_0^1\ol y(x,\l_o)\big(-y''(x,\l_o)+(V(x,\l_o)-\l_o)y(x,\l_o)\big)dx
\\
=\int_0^1\big(|y'(x,\l_o)|^2+(V(x,\l_o)-\l_o)|y(x,\l_o)|^2\big)dx,
\end{aligned}
$$
which yields $\int_0^1\eta(x,\l_o)|y(x,\l_o)|^2dx=0$.
This identity shows that $\eta(x,\l_o)$
vanishes at least at one point
in the interval $x\in(0,1)$.

The definition \er{defb} gives
$$
|\eta(x,\l)|\ge \big||\nu|-|Q(x,\l)|\big|\qqq\forall\ \ (x,\l)\in\R\ts\mD.
$$
If $|\nu|>\sup_{x\in[0,1]}|Q(x,\l)|$, then $|\eta(x,\l)|>0$
for all $x\in[0,1]$.
Therefore, $\eta(x,\l)$
may vanish only if $|\nu|\le\sup_{x\in[0,1]}|Q(x,\l)|$.
This yields the estimate \er{qest}.
\BBox

\medskip

We prove our first main results about the spectra.

\medskip

\no {\bf Proof of Theorem~\ref{ThSpec2}.}
Let $x\in\R$. The function $ V(x,\cdot)$ is real analytic in $\mD$
and $Q(x,\cdot)\in C(\ol\mD)$,
then $\eta(x,\cdot)$ is harmonic in $\mD$, each
$\eta(x,\cdot),\pa_\nu\eta(x,\cdot)\in C(\ol\mD)$, and
$\eta(x,\mu)=0$ for all $\mu\in I$.
Moreover,
$$
\pa_\nu\eta(x,\l)=\pa_\nu Q(x,\l)-1,\qqq \pa_\nu={\pa\/\pa\nu}
$$
which yields
$$
\big|\pa_\nu\eta(x,\l)\big|\ge\big|1-|\pa_\nu Q(x,\l)|\big|.
$$
for all $(x,\l)\in\R\ts\mD$.
The estimate \er{locestdil} implies
$
\sup_{x\in[0,1]}\big|\pa_\nu Q(x,\mu+i\nu)|_{\nu=0}\big|<1
$,
which yields
$
\inf_{x\in[0,1]}\big|\pa_\nu\eta(x,\mu+i\nu)|_{\nu=0}\big|> 0
$
for all $\mu\in I$.

Thus, we have $\eta(x,\mu)=0$ for all $(x,\mu)\in \R\ts I$, and
$
\inf_{x\in[0,1]}\big|\pa_\nu\eta(x,\mu+i\nu)|_{\nu=0}\big|> 0
$
for all $\mu\in I$. Then the asymptotics
$$
\e(x,\mu+i\nu)=\pa_\nu\eta(x,\mu+i\nu)|_{\nu=0}\nu+O(\nu^2),\qqq \nu\to 0,
$$
yields
$|\e(x,\mu+i\nu)|>0$ for all $x\in\R$, $\mu\in I$ and $|\nu|<\d$ for some $\d$ small enough.
Then Lemma~\ref{LmVis0} shows that there are no the spectra
$\s(H(k,\cdot))$ for each $k\in[0,2\pi)$ and the spectrum $\s(T)$
in the domain $(\mu,\nu)\in I\ts(-\d,\d)$, which yields \er{locrsp}.
\BBox

\subsection{Auxiliary estimate}
Below we search for the conditions for the potential, when
the high energy spectrum is real.
In our proofs we use the arguments from
\cite{McK81}.

Below we need the following auxiliary result.

\begin{lemma}
\lb{Lmehf}
Let the function $f$ be harmonic in the disc
$\dD_{\mu}(r )=\{\l\in\C:|\l-\mu|<r\}$ for some $r>0, \mu\in\R$.
Then
\[
\lb{estda}
\Big|{\pa f(\mu+i\nu)\/\pa\nu}\Big|\le{2 r\/( r-|\nu|)^2}
\max_{\l\in \dD_{\mu}(r )}|f(\l)|,\qq \forall\ \ \nu\in(-r,r).
\]
If, in addition,  $-\phi r\le\nu\le\phi r$
for some $\phi\in(0,1)$  and
\[
\lb{esta}
\max_{\l\in \dD_{\mu}(r )}|f(\l)|<{r(1-\phi)^2\/2},
\]
then
\[
\lb{estdera}
\Big|{\pa f(\mu+i\nu)\/\pa\nu}\Big|<1.
\]
\end{lemma}

\no {\bf Proof.}
Consider the case $\nu>0$. Poisson's formula for the disc $\dD_{\mu}(r )$ gives
$$
f(\mu+i\nu)={1\/2\pi}\int_0^{2\pi}
{(r^2-\nu^2)f(\mu+r e^{i\theta})\/r^2+\nu^2-2r \nu\sin\theta}d\theta,
$$
which yields
\[
\lb{Puasfd}
{\pa f(\mu+i\nu)\/\pa\nu}={r\/\pi}\int_0^{2\pi}
{\nu^2\sin\theta+r^2\sin\theta-2\nu r\/(r^2+\nu^2-2r \nu\sin\theta)^2}
f(\mu+r e^{i\theta})d\theta.
\]
The estimates
$$
\begin{aligned}
&r^2+\nu^2-2r \nu\sin\theta=(r-\nu)^2+2r\nu(1-\sin\theta)\ge(r-\nu)^2,\\
&\nu^2\sin\theta+r^2\sin\theta-2\nu r\le(\nu- r)^2
\end{aligned}
$$
give
$$
\Big|{\pa f(\mu+i\nu)\/\pa\nu}\Big|\le{2 r\/( r-\nu)^2}
\max_{\theta\in[0,2\pi]}|f(\mu+ r e^{i\theta})|,
$$
which yields \er{estda} for the case $\nu>0$.
The arguments for the case $\nu<0$ are similar.
Let $\nu\to 0$ in the identity \er{Puasfd}, then we obtain
$$
{\pa f(\mu+i\nu)\/\pa\nu}\Big|_{\nu=0}={1\/\pi r}\int_0^{2\pi}
f(\mu+ r e^{i\theta})\sin\theta d\theta,
$$
which yields \er{estda} for the case $\nu=0$.
The estimate \er{estda} gives \er{estdera}.
\BBox

\medskip

\subsection{Global conditions}
Introduce the domains in $\C$:
$$
\Pi_{a,b}(r)=(a,b)\ts(-r,r),\qq
\Pi_{a}(r)=(a,+\iy)\ts(-r,r), \qq a,b\in\R,\qq a<b,\qq r>0.
$$
Now we prove that the spectra are real under some specific restriction on the potential.

\begin{lemma}
\lb{Lmlsp}
Let $\Pi_{a}( r )\ss\mD$  for some $(a, r)\in\R\ts\R_+$.

i) Let, in addition, $b>a+2 r$ and let $|Q(x,\l)|$
be bounded in $[0,1]\ts \Pi_{a,b}( r)$ and satisfy
\[
\lb{estQ}
\xi(\Pi_{a,b}( r))\le{ r(1-\phi)^2\/2}
\]
for some $\phi\in(0,1)$, where the functional $\xi$ is given by \er{defunc}.
Then for all $(x,\l)\in\R\ts\Pi_{a+ r,b- r}(\phi r)$
the function $\eta(x,\l)=Q(x,\l)-\nu$ can vanish only for real $\l$.
Moreover, the spectra $\s(H)$,  $\s(T)$ and $\s(\cN)$
in the rectangle $\Pi_{a+ r,b- r}(\phi r)$ are real:
\[
\lb{locspr}
\gS\cap\Pi_{a+ r,b- r}(\phi r)\ss\R.
\]

ii) Let, in addition, $|Q(x,\l)|$
be bounded in $[0,1]\ts \Pi_{a}( r)$
and satisfy
\[
\lb{estVhs}
\xi(\Pi_{a}( r))\le{ r(1-\phi)^2\/2}
\]
for some $\phi\in(0,1)$.
Then
the spectra in the half strip domain $\Pi_{a+ r}(\phi r)$ are real:
\[
\lb{locsprhs}
\gS\cap\Pi_{a+ r}(\phi r)\ss(a+ r,+\iy).
\]
Moreover, if
\[
\lb{estxi23}
\xi\big(\Pi_{a}( r)\big)\le(2-\sqrt3)r,
\]
then
\[
\lb{incl1}
\gS\cap\Pi_{a+ r}\big((2-\sqrt3) r\big)\ss(a+ r,+\iy).
\]

\end{lemma}

\no {\bf Proof.}
i) Let $x\in\R$. Due to  $ V(x,\l)$ is real for $\l\in(a,b)$, we have
$\eta(x,\l)=0$
as $\l\in(a,b)$.  Let, in addition, $\mu\in(a+ r,b- r)$.
The function $Q(x,\cdot)$ is harmonic in $\Pi_{a,b}( r)$ and satisfies \er{estQ},
then the estimate \er{estdera} shows that
$$
\Big|{\pa Q(x,\mu+i\nu)\/\pa\nu}\Big|<1,\qqq \forall\qq |\nu|\le\phi r,
$$
which yields
$$
{\pa \eta(x,\mu+i\nu)\/\pa\nu}>0\qq\text{ or}\qq {\pa \eta(x,\mu+i\nu)\/\pa\nu}<0
,\qqq \forall\qq |\nu|\le\phi r.
$$
Consider the case ${\pa \eta/\pa\nu}>0$.
Then $\eta(x,\mu+i\nu)>0$, if
$\nu\in(0,\phi r)$, and $\eta(x,\mu+i\nu)<0$, if
$\nu\in(-\phi r,0)$.
The similar arguments for the case ${\pa \eta/\pa\nu}<0$ hold.
Thus for all $x\in\R$ and for all $\l\in\Pi_{a+ r,b- r}(\phi r)$
the function $\eta(x,\l)$ can vanish only for real $\l$.

Let $\l_o\in\gS$. Lemma~\ref{LmVis0}~i) yields that
$\eta(x_o,\l_o)=0$ for some $x_o\in(0,1)$.
If, in addition, $\l_o\in\Pi_{a+ r,b- r}(\phi r)$, then
the statement i) implies $\l_o\in\R$. The relation \er{locspr} follows.

ii) Taking $b\to+\iy$ in \er{estQ} and \er{locspr} we obtain \er{estVhs} and \er{locsprhs}.
If $\phi=2-\sqrt3$, then $(1-\phi)^2=2\phi$.
The relations \er{estVhs} and \er{locsprhs} imply \er{estxi23} and \er{incl1}.
\BBox

\medskip

We are ready to prove Theorem~\ref{Thspec}.
%The proof of Theorem~\ref{Thspec} is based on the following considerations.
%Let $\Pi_{a}( r )\ss\mD$  for some $(a, r)\in\R\ts\R_+$.
%Lemma~\ref{Lmlsp}~ii) shows that if the estimate \er{estVhs} is fulfilled
%%\[
%%\lb{ests1}
%%\xi(\Pi_{a}( r))\le{ r(1-\phi)^2\/2}
%%\]
%for some $\phi\in(0,1)$, then
%the spectra in the half strip domain $\Pi_{a+ r}(\phi r)$ are real.
%The estimate \er{qest} gives that
%all spectra in the domain $\mD_{a+r}$ lie in the strip
%$
%\Pi_{a+r}(\beta),$ where
%$$
%\beta=\xi(\mD_{a+r}).
%$$
%Assume that
%\[
%\lb{ests2}
%\xi(\mD_{a})\le \phi r\qq
%\text{and}\qq \phi\le{(1-\phi)^2\/2}.
%\]
%Then:
%
%1) \er{estVhs} is fulfilled,
%
%2) $\beta\le \phi r$, therefore, $\Pi_{a+r}(\beta)\ss\Pi_{a+ r}(\phi r)$.
%
%\no The item 1) yields that the spectra in the domain $\Pi_{a+ r}(\phi r)$ are real.
%Then the item 2) shows that all spectra in the domain $\mD_{a+r}$ are real.
%The strongest result is obtained if we take $\phi$ in \er{ests2}
%such that $\phi={(1-\phi)^2\/2}$.

\medskip

\no {\bf Proof of Theorem~\ref{Thspec}.}
%Note that $\phi=2-\sqrt3$
%satisfies the identity $(1-\phi)^2=2\phi$.
The definition \er{cestimV2}
gives
%that $Q(x,\l)$ satisfies the estimate
$$
\xi(\mD_{a+\rho})\le
\xi(\mD_{a})=\phi \rho,\qqq \phi=2-\sqrt3.
$$
Then the estimate \er{qest} shows that if $\l_o=\mu_o+i\nu_o\in\gS\cap\mD_{a+\rho}$,
then
$$
|\nu_o|\le \xi(\mD_{a+\rho})\le\phi \rho,
$$
therefore, $\l_o\in\Pi_{a+\rho}(\phi \rho)$. This yields
\[
\lb{incl2}
\gS\cap\mD_{a+\rho}\ss\Pi_{a+\rho}(\phi \rho).
\]
The relations \er{incl1} and \er{incl2} give \er{locsphs},
which yields \er{locsphs1}.
\BBox

\subsection{Examples}
\lb{SSex}
The following examples illustrate Theorems~\ref{ThSpec2} and \ref{Thspec}.
We consider the potentials $V$ of the forms
\[
\lb{ex1pot}
V_1(x,\l)=\sum_{n=1}^Nq_n(x)e^{-\kappa_n\l},
\]
\[
\lb{ex2pot}
V_2(x,\l)=\sum_{n=1}^Nq_n(x)\cos(\kappa_n\l),
\]
for all $(x,\l)\in\R\ts\C$, where
%Let  $\kappa_n$ be positive numbers and  $q_n$ be real functions such that
\[
\lb{excaq}
0<\kappa_1<\kappa_2<...<\kappa_N,\qq q_n\in L_{real}^\iy(\T),\qq n=1,...,N,\qq N\ge 1.
\]
Introduce the norm $\|f\|_\iy=\sup_{x\in[0,1]}|f(x)|$.

\begin{proposition}
Let  $\kappa_n,n=1,...,N,N\in\N$, be positive numbers and
let $q_n$ be real functions satisfying \er{excaq}.

i) If $V=V_1$, then

a) The spectra $\s(H)$, $\s(T)$ and $\s(\cN)$
in the strip $\R\ts(-\d,\d)$ for some $\d>0$ are real:
\[
\lb{ex1is}
\gS\cap(\R\ts(-\d,\d))\ss\R.
\]

b) The spectra $\s(H)$, $\s(T)$ and $\s(\cN)$ in the half-plane $\Re\l>\mu_1$,
are real:
\[
\lb{ex1ihp}
\gS\cap\Pi_{\mu_1}\ss\R,
\]
where
$$
\mu_1={1\/2-\sqrt3}\sum_{n=1}^N\|q_n\|_\iy.
$$

ii)  If $V=V_2$,
then

a) The spectra $\s(H)$, $\s(T)$ and $\s(\cN)$
in the strip $\R\ts(-\d,\d)$ for some $\d>0$ are real:
\[
\lb{ex2is}
\gS\cap(\R\ts(-\d,\d))\ss\R.
\]
%the spectra $\s(H)$ and $\s(T)$
%in the strip $\R\ts(-\d,\d)$ for some $\d>0$ are real.

b) For any $\nu_0>0$ the spectra $\s(H)$, $\s(T)$ and $\s(\cN)$
in the half-strip $(\mu_0,+\iy)\ts(-\nu_0,\nu_0)$ are real:
\[
\lb{ex2ihs}
\gS\cap\Pi_{\mu_0}(\nu_0)\ss\R,
\]
where
$$
\mu_0={1\/2-\sqrt3}\sum_{n=1}^N\|q_n\|_\iy\sinh(\kappa_n\nu_0).
$$

\end{proposition}

\no {\bf Proof.}
i) Let the potential have the form \er{ex1pot}.
Then
$$
Q(x,\l)=-\sum_{n=1}^N q_n(x)e^{-\kappa_n\mu}\sin(\kappa_n\nu),\qq \l=\mu+i\nu
$$
which yields
$$
\sup_{(x,\l)\in [0,1]\ts \R}\Big|{\pa Q(x,\l)\/\pa\nu}\Big|\le
\sum_{n=1}^N\kappa_n\|q_n\|_\iy,
$$
$$
\xi(\Pi_{0})\le\sum_{n=1}^N\|q_n\|_\iy.
$$
Theorem~\ref{ThSpec2} gives \er{ex1is}.
%shows that if $\sum_{n=1}^N\kappa_n\|q_n\|_\iy<1$, then
%the spectra $\s(H)$ and $\s(T)$
%in the strip $\R\ts(-\d,\d)$ for some $\d>0$ are real.
Theorem~\ref{Thspec} yields \er{ex1ihp}.
%that the spectra
%in the half-plane
%$$
%\Re\l>{1\/2-\sqrt3}\sum_{n=1}^N\|q_n\|_\iy
%$$
%are real.

ii) Let the potential have the form \er{ex2pot}.
Then
$$
Q(x,\l)=-\sum_{n=1}^N q_n(x)\sin(\kappa_n\mu)\sinh(\kappa_n\nu).
$$
which yields
$$
\sup_{(x,\l)\in [0,1]\ts \R}\Big|{\pa Q(x,\l)\/\pa\nu}\Big|\le
\sum_{n=1}^N\kappa_n\|q_n\|_\iy.
$$
Theorem~\ref{ThSpec2} implies \er{ex2is}.

Consider  the half-strip domain $\Pi_{0}(\nu_0),\nu_0>0$. Then
$$
\xi(\Pi_{0}(\nu_0))\le\sum_{n=1}^N\|q_n\|_\iy\sinh(\kappa_n\nu_0).
$$
Theorem~\ref{Thspec} yields \er{ex2ihs}.
\BBox

\section{High energy spectrum}
\setcounter{equation}{0}

\subsection{Spectral properties}

Theorem~\ref{ThSpec2} shows that if $\Pi_{a}( r )=(a,+\iy)\ts(-r,r)\ss\mD$
for some $(a, r)\in\R\ts\R_+$
and the potential $ V$ satisfies the condition \er{estdQhl},
then the spectrum $\s(H)$ on the half-strip
$\Pi_a(\d)$ for some  $\d>0$ is real.
Using this result and the standard arguments
based on the analyticity of the Lyapunov function $\D(\l)$, see \cite{Kr83},
we obtain the following results about this function.

\begin{lemma}
\lb{LmpLf}
Let $\Pi_{a}( r )\ss\mD$ for some $(a, r)\in\R\ts\R_+$, let
$Q=\Im V$ satisfy the estimate \er{estdQhl} and let  $\l\in(a,+\iy)$.
Then if $\D(\l)\in(-1,1)$, then
$\D'(\l)\ne 0$.
Moreover, if $\D(\l)=\pm 1$ and $\D'(\l)=0$, then $\D(\l)\D''(\l)<0$.

\end{lemma}

\no {\bf Proof.}
We give the proof by the method of ``on the contrary''.
Assume that $\l_o\in(a,+\iy)$ satisfies $\D(\l_o)\in(-1,1)$
and $\D'(\l_o)=0$.
Then $\D(\l)=\D(\l_o)+{1\/2}\D''(\l_o)(\l-\l_o)^2+O((\l-\l_o)^3)$
as $\l-\l_o\to 0$.
Consider the mapping $\l\to \D(\l)$ in some neighborhood of the point
$\l_o$. Any angle made by lines started from the point $\l_o$,
is transformed onto the angle two or more times grater.
Then the segment
$[\D(\l_o)-\d,\D(\l_o)+\d]\ss[-1,1]$ for some $\d>0$ small enough has the pre-image,
that cannot entirely lie on the real axis.
The identity \er{splf} gives that $\s(H(k,\cdot))$ is non-real for some $k\in[0,2\pi)$.
The identity \er{specH} implies that $\s(H)$ is non-real
that contradicts to Theorem~\ref{ThSpec2}.
Thus, $\D'(\l_o)\ne 0$, which proves the first statement.
The proof of the second one is similar.
\BBox

\medskip

In the unperturbed case $ V=0$ the spectrum $\s(H_o(k)),k\in[0,\pi]$, consists of
the eigenvalues
\[
\lb{lnok}
\l_{2n+1}^o(k)=(2\pi n+ k)^2,\ \ n=0,1,2,...,\qq\l_{2n}^o(k)=(2\pi n- k)^2,\ \ n\in\N,
\]
$$
\l_1^o(k)\le\l_2^o(k)\le\l_3^o(k)\le\l_4^o(k)<...
$$
If $k\in(0,\pi)$, then all eigenvalues are simple.
If $k=0$, then $\l_1^o(0)$ is simple and all other eigenvalues has
multiplicity $2$. If $k=\pi$, then all eigenvalues have
multiplicity $2$.
Moreover, using $\s(H(2\pi-k,\cdot))=\s(H(k,\cdot))$ for all $k\in[0,2\pi)$,
we put $\l_n^o(2\pi-k)=\l_n^o(k),n\in\N$.

%Now we describe the high energy spectrum of the quasi-periodic problem.

\begin{lemma}
\lb{Lm2pev}
Let $\Pi_{a}( r )\ss\mD$ for some $(a, r)\in\R\ts\R_+$.
Let the potential $ V$ satisfy the condition \er{estdQhl}
and let $\| V(\cdot,\l)\|=z o(1)$
as $\l\to+\iy$, $z=\l^{1\/2}>0$.
Let $b> a$ be large enough.
Then

i) There exist
exactly two (counting with multiplicity) eigenvalues $\l_{2n}^\pm$
of the operator $H(0,\l)$ in each interval $(((2n-1)\pi )^2,((2n+1)\pi )^2)\ss\mD_b,n\in\N$,
and exactly two (counting with multiplicity) eigenvalues $\l_{2n-1}^\pm$
of the operator $H(\pi,\l)$ in each interval $(((2n-2)\pi )^2,(2n\pi )^2)\ss\mD_b,n\in\N$,
and there are no other eigenvalues in $\mD_{b}$.

ii)
The eigenvalues $\l_n(k)\in\mD_b,k\in(0,\pi)$,
of the operator $H(k,\l)$ are simple and satisfy
\[
\lb{indp2}
\l_{N}(k)<\l_{N+1}(k)<\l_{N+2}(k)<\l_{2N+1}(k)<...,
\]
for some $N\in\N$. Moreover,
\[
\lb{indp3}
\l_{2n}(k)\in(\l_{4n-3}^+,\l_{4n-2}^-),\qq \tf{d\l_{2n}(k)}{dk}<0,\qq
\l_{2n}(0)=\l_{4n-2}^-,\qq  \l_{2n}(\pi)=\l_{4n-3}^+,
\]
\[
\lb{indp4}
\l_{2n-1}(k)\in[\l_{4n-4}^+,\l_{4n-3}^-],\qq \tf{d\l_{2n-1}(k)}{dk}>0,\qq
\l_{2n-1}(0)=\l_{4n-4}^+,\qq \l_{2n-1}(\pi)=\l_{4n-3}^-,
\]
where $n=\tf{N+1}{2},\tf{N+3}{2},...$,
and recall $\l_n(k)=\l_n(2\pi-k)$.
There are no other eigenvalues of the operator $H(k,\l)$ in $\mD_b$.

iii) There exists
exactly one simple eigenvalue $\gm_n$
of the Dirichlet operator $T(\l)$
and exactly one simple eigenvalue $\gn_n$
of the Neumann operator $\cN(\l)$
 in each interval
$((\pi(n-{1\/2})^2,(\pi(n+{1\/2})^2))\ss\mD_b,n\in\N$,
and there are no other eigenvalues in $\mD_{b}$.

iv) The Lyapunov function  satisfies
\[
\lb{lfgb}
\begin{aligned}
-1<\D(\l)<1,&\qq\as\qq \l\in(\l_{2n-1}^+,\l_{2n}^-),\\
\D(\l)>1,&\qq\as\qq \l\in(\l_{2n}^-,\l_{2n}^+),\\
\D(\l)<-1,&\qq\as\qq \l\in(\l_{2n-1}^-,\l_{2n-1}^+),
\end{aligned}
\]
for all $n>N$, where $N\in\N$ is large enough.

v) The relations \er{dsing} hold true.
\end{lemma}

\no {\bf Proof.}
i) We consider the eigenvalues of the periodic problem,
the proof for  the anti-periodic ones is similar.
We proved in Theorem~\ref{Thspec} that the eigenvalues of the periodic
problem in $\mD_{b}$
are real and the identity \er{speck2p} shows that they are zeros of
the function $\D(\l)-1$. The estimate \er{estLfr} implies
\[
\lb{asLfr}
\D(\l)=\cos z+o(1),\qq
\D(\l)-1=-\sin^2{z\/2}+o(1),\qq  \l\to+\iy.
\]
This asymptotics shows that
for each $n\in\N$ large enough there exists
exactly two (counting with multiplicity) eigenvalue
of the periodic problem in the interval $(((2n-1)\pi )^2,((2n+1)\pi )^2)$
and there are no other eigenvalues in $\mD_{b}$.

ii) Lemma~\ref{LmpLf} and the asymptotics \er{asLfsp} show that the Lyapunov function
at high energies behaves in the similar way as in the case
of the Schrodinger operator with a potential, which does not depend on energy.
Exactly, it oscillates as follows. It increases from -1 to 1,
then either immediately starts to decrease, or first it becomes more than 1,
and then it goes back to the value 1.
After that, it decreases from 1 to -1, then it
either immediately starts to increase, or first it becomes less than -1,
and then returns back to -1.
Further, the process is repeated again and again to infinity.
Thus, the zeros $\l_n(k)$ of the function
$\D-\cos k$ (the eigenvalues of the problem \er{2oequd}, \er{fbc}) at high energy satisfy
\er{indp2}--\er{indp4}.

iii) We have proved in Theorem~\ref{Thspec} that the eigenvalues of the Dirichlet
problem in $\mD_{b}$
are real and the identity \er{specd} shows that they are zeros of
the function $\vp(1,\l)=0$.
The estimate \er{estfir} gives
$\vp(1,\l)={1\/z}(\sin z+o(1))$ as $\l\to+\iy.$
This asymptotics shows that
for each $n\in\N$ large enough there exists
exactly one simple eigenvalue
of the Dirichlet problem in the interval $((\pi(n-{1\/2})^2,(\pi(n+{1\/2})^2)$
and there are no other eigenvalues in $\mD_{b}$.
The proof for the Neumann operator is similar.

iv) The identities \er{splf} and \er{speck2p},
Lemma~\ref{LmpLf} and the asymptotics \er{asLfr}
imply \er{lfgb}.

v) We have the identities
\[
\lb{D^2}
\begin{aligned}
\D(\l)^2=\Big({\vt(1,\l)+\vp'(1,\l)\/2}\Big)^2
=\Big({\vt(1,\l)-\vp'(1,\l)\/2}\Big)^2+\vt(1,\l)\vp'(1,\l)
\\
=\Big({\vt(1,\l)-\vp'(1,\l)\/2}\Big)^2+\vt'(1,\l)\vp(1,\l)+1
\end{aligned}
\]
for all $\l\in\mD$. Let $\l\in\s(T)$. Then $\vp(1,\l)=0$
and we obtain $\D^2(\l)\ge 1$. The estimates \er{lfgb} give \er{dsing}.
The proof for the Neumann eigenvalues is similar.
\BBox

\medskip

\no {\bf Remark.} Similarly we can consider the mix problems $y(0)=y'(1)=0$
and $y'(0)=y(1)=0$
for equation \er{2oequd}. The spectra are discrete
and the large eigenvalues are simple and belong to the intervals
$
[\l_{n-1}^+,\l_{n}^-],n=N,N+1,...
$

\subsection{Spectral asymptotics}
Now we determine high energy eigenvalue asymptotics
for the operator $H(k,\l)$.

\begin{proposition}
Let $\Pi_{a}( r )\ss\mD$ for some $(a, r)\in\R\ts\R_+$.
Let the potential $ V$ satisfy the conditions \er{estdQhl} and \er{condVav}
and let $\| V(\cdot,\l)\|=z^{1\/2}o(1)$ as $\l\to+\iy$.
Then the eigenvalues of the operator $H(k)$ satisfy
\[
\lb{asqphe}
\l_n(k)=\l_n^o(k)+o(1),\qqq\forall\ \ k\in(0,\pi),
\]
as $n\to+\iy$, where $\l_n^o(k)$ are given by \er{lnok}.
\end{proposition}

\no {\bf Proof.}
Let $0<k<\pi$.
We prove the asymptotics \er{asqphe} for $\l_{2n+1}(k)$,
the proof for $\l_{2n}(k)$ is similar.
Let $\l=\l_{2n+1}(k)$ for some $n\in\N$ large enough.
Then $z=\l^{1\/2}=2\pi n+k+\d,\d=\d_n=O(1)$, as $n\to+\iy$,
and the estimate \er{asLfm} gives
\[
\lb{asDqpev}
\D(\l)-\cos k=-2\sin{ \d\/2}\sin\Big(k+{\d\/2}\Big)
+{\sin (k+\d)\/4\pi n}\wh V_o(\l)+{o(1)\/n}.
\]
The identity $\D(\l)-\cos k=0$ gives $\d=O(n^{-1})$ and using \er{condVav} and
\er{asDqpev} again we obtain
$$
\D(\l)-\cos k=-2\sin{ \d\/2}\sin k+{o(1)\/n}.
%\qq\l_o=\l_{2n+1}^o(k).
$$
Now the identity $\D(\l)-\cos k=0$ gives
$
\d=o(n^{-1}).
$
Then
$
z=2\pi n+k+o(n^{-1}),
$
which yields the asymptotics \er{asqphe}.
\BBox

\medskip

Now we prove our  results about the high energy asymptotics of the spectra
of the operator $H(\l)$.

\medskip

\no {\bf Proof of Theorem~\ref{Thasspec}.}
i) Due to Theorem~\ref{ThSpec2}, the spectra are real.
Lemma~\ref{Lm2pev}~i) yields \er{indp1}.
Lemma~\ref{Lm2pev}~ii) and the identity \er{specH} give \er{specHreal}.
The relation \er{dsing} is proved in Lemma~\ref{Lm2pev}~v).

ii) Let $\l=\l_{2n}^\pm$.
Then $z=\l^{1\/2}=2\pi n+\d,\d=\d_n=O(1)$, as $n\to+\iy$,
and the estimate \er{asLfsp} gives
\[
\lb{asDpev}
\D(\l)-1=-2\sin^2{ \d\/2}+{\sin \d\/4\pi n}\wh V_o(\l)+{O(1)\/n^2}
=-2\Big(\sin{ \d\/2}-{\cos{\d\/2}\/8\pi n}\wh V_o(\l)\Big)^2+{O(1)\/n^2}.
\]
The identity $\D(\l)-1=0$ implies $\d=O(n^{-1})$.
Using the asymptotics \er{asLfsp} again we obtain
$$
\D(\l)-1=-{\d^2\/2}+{\d\/4\pi n}\wh V_o(\l)
+o(n^{-2})
=-{1\/2}\Big(\d-{\wh V_o(\l)\/4\pi n}\Big)^2
+o(n^{-2}).
$$
The identity $\D(\l)-1=0$ and the condition \er{condVav} give
$
\d=o(n^{-1}),
$
which yields \er{as2p} for $n$ even. The proof for $n$ odd is similar.
\BBox

\medskip

Now we prove the results about the good Boussinesq equation.

\medskip

\no {\bf Proof of Corollary~\ref{Corbe}.}
Let $p',q\in L^1(\T)$. Then the solution $\p_3$ of equation \er{1b}
satisfies $\p_3'''(\cdot,\zeta)\in L^1(\T)$ for all  $\zeta\in\cD$.
The definition \er{da1} and the asymptotics \er{asa} show that the function $V$,
given by \er{VBe}, satisfies:
$\|V(\cdot,\l)\|$ is uniformly bounded in $\mD$
and  $|\Im V(x,\zeta)|$ is uniformly bounded in $[0,1]\ts\mD$, where the domain $\mD$
has the form \er{mDBe}.
 The relation \er{locsphs1} yields that
the ramifications $r_n^\pm$ and the three-point eigenvalues $\zeta_n$ in the
half-plane $\cZ_a$ are real. Lemma~\ref{Lm2pev} gives that
there are exactly two ramifications $r_{n}^\pm$
and exactly one simple eigenvalue $\zeta_n$
in each interval $(\a_n^-,\a_n^+)$
inside this half-plane.
Moreover, the estimate \er{locestdil} holds true for all
$\l>0$ large enough (see Remarks to Theorem~\ref{Thspec}).
Then the asymptotics \er{as2p} implies
\er{asram}. The relations \er{dsing}
give \er{as3p}.
\BBox

\medskip

\no\small {\bf Acknowledgments.}
A.~Badanin was supported by the RFBR grant number 19-01-00094.
E.~Korotyaev was supported by the RSF grant number 18-11-00032.


\begin{thebibliography}{9999}
\setlength{\itemsep}{-\parskip}
\footnotesize

\bibitem[A80]{A80}  L. M.Alonso, Schr\"odinger spectral problems
with energy--dependent potentials as sources of nonlinear
Hamiltonian evolution equations, Journal of Mathematical Physics,
21(9) (1980), 2342--2349.

\bibitem[BK11]{BK11} A.Badanin,  E.Korotyaev,
Spectral estimates for periodic fourth order operators,
St.Petersburg Math. J. 22:5 (2011) 703--736.

\bibitem[BK12]{BK12} A.Badanin, E. L.Korotyaev,
Even order periodic operators on the real line,
International Mathematics Research Notices 2012(5) (2012)
1143--1194.

\bibitem[BK14]{BK14} A.Badanin,  E.Korotyaev,
Third order operator with periodic coefficients on the real axis,
St. Petersburg Math. J. 25:5 (2014) 713--734.


\bibitem[BK15]{BK15} A.Badanin, E.Korotyaev. Spectral asymptotics for
the third order operator with periodic coefficients.
Journal of Differential Equations 253 (2012) 3113--3146.

\bibitem[BK19]{BK19} A.Badanin, E.Korotyaev,
Third-order operators with three-point conditions associated
with Boussinesq's equation, Applicable Analysis (2019),
DOI: 10.1080/00036811.2019.1610941.

\bibitem[CJ67]{CJ67} F. Calogero, G. Jagannathan,
Levinson's theorem for energy-dependent potentials,
Il Nuovo Cimento A (1965-1970) 47(2) (1967) 178--188.

\bibitem[DM89]{DM89} V.A.Derkach,   M.M.Malamud,
Some classes of analytic operator-valued functions with a nonnegative imaginary
part. (Russian. English summary)
Dokl. Akad. Nauk Ukrain. SSR Ser. A (1989) no. 3, 13--17, 87.

\bibitem[FLM04]{FLM04} J. Form\'anek, R. J. Lombard, J. Mare\v{s},
Wave equations with energy-dependent potentials,
Czechoslovak journal of physics, 54(3) (2004) 289--315.

\bibitem[GKMT01]{GKMT01} F.Gesztesy,  N.J.Kalton,  K.A.Makarov,   E.Tsekanovskii,
Some applications of operator-valued Herglotz functions,
In Operator theory, system theory and related topics,
Birkh\"auser, Basel, 2001, 271--321.

\bibitem[JJ76]{JJ76} M.Jaulent, C.Jean,  The inverse problem
for the one-dimensional Schr\"odinger
equation with an energy-dependent potential, I. Ann. Inst. Henri Poincar\'e
25(2) (1976) 105--118.

\bibitem[JJ76x]{JJ76x} M.Jaulent, C.Jean, The inverse problem for
the one-dimensional Schr\"odinger equation with an energy-dependent potential. II,
Ann. Inst. Henri Poincar\'e 25(2) (1976) 119--137.


\bibitem[Ka08]{Ka08} Y.Kamimura, Energy dependent inverse scattering on the line,
Differential and Integral Equations 21(11-12) (2008) 1083--1112.

\bibitem[Ke71]{Ke71}  M.V.Keldysh, On the completeness of the eigenfunctions
of some classes of non-selfadjoint linear operators,
 Russian Math. Surveys 26:4 (1971) 15--44.

\bibitem[K99]{K99}  E.Korotyaev, Inverse Problem and
the trace formula for the Hill Operator, II,
Mathematische Zeitschrift 231(2) (1999) 345--368.

\bibitem[K03]{K03} E.Korotyaev, Characterization of the spectrum of
Schr\"odinger operators with periodic distributions, Int. Math.
Res. Not.  (2003) no. 37 2019--2031.

\bibitem[Kr83]{Kr83} M.G.Krein,  The basic propositions of the theory
of $\l$-zones of stability of a canonical system of linear
differential equations with periodic coefficients, In Topics in Differential
and Integral Equations and Operator Theory (pp. 1-105), (1983), Birkh\"auser, Basel.


\bibitem[Ma12]{Ma12} A.S.Markus,  Introduction
to the spectral theory of polynomial operator pencils,
American Mathematical Soc. (2012).


\bibitem[McK81]{McK81} H.McKean, Boussinesq's equation on the circle,
Com. Pure and Appl. Math. 34(1981) 599--691.


\bibitem[P95]{P95}  V.Papanicolaou,
The spectral theory of the vibrating periodic beam,
Commun. Math. Phys. 170 (1995) 359 -- 373.

\bibitem[P03]{P03}  V.Papanicolaou,
The periodic Euler-Bernoulli equation,
Trans. Amer. Math. Soc. 355 (2003), no. 9, 3727--3759.

\bibitem[PT87]{PT87} J.P\"oschel, E.Trubowitz, Inverse spectral
theory, Academic Press, Boston, 1987.

\bibitem[RS78]{RS78} M.Reed,   B.Simon,  Methods of modern mathematical
physics. IV. Analysis of operators, Academic Press, New York-London,
1978.



\end{thebibliography}
\end{document}